\documentclass[12pt]{article}
\pdfoutput=1

\usepackage{url}
\usepackage{setspace}
\usepackage{amsmath,amssymb,amscd,amsthm}
\usepackage{bbold}
\usepackage{amsfonts}
\usepackage{color}
\usepackage{graphicx}
\usepackage{hyperref}
\usepackage{cleveref}
\usepackage{subcaption}
\usepackage{cite}

\newcommand{\bea}{\begin{eqnarray}}
\newcommand{\eea}{\end{eqnarray}}
\newcommand{\be}{\begin{equation}}
\newcommand{\ee}{\end{equation}}
\newcommand{\ba}{\begin{align}}
\newcommand{\ea}{\end{align}}

\newcommand{\CC}{\mathbb{C}} % Complex
\newcommand{\HH}{\mathbb{H}} % Hyperbolic space
\newcommand{\QQ}{\mathbb{Q}} % Rationals
\newcommand{\ZZ}{\mathbb{Z}} % Integers
\newcommand{\id}{\mathbb{1}} % Identity

\newcommand{\op}{\mathcal{O}}

\DeclareMathOperator{\Tr}{Tr}
\newcommand\rref[1]{(\ref{#1})}
\newcommand{\F}{{F}}
\renewcommand{\Re}{\operatorname{Re}}
\renewcommand{\Im}{\operatorname{Im}}

\newcommand{\xbar}{\bar{x}}
\newcommand{\calF}{\mathcal{F}}

\newcommand{\twoFone}{{{}_2 F_1}}

\newcommand{\rats}{\mathcal{R}} % Set of rational tangles

\begin{document}

\title{\vspace{-1cm}\begin{flushright}\end{flushright}\vspace{2cm}\bf{A conformal block Farey tail}}

\author{Alexander Maloney, Henry Maxfield, Gim Seng Ng}
\maketitle
\begin{center}
{\it  Department of Physics, McGill University, Montr\'eal, Canada }
	 \smallskip

\end{center}
\vspace{3em}

\begin{center}
{\bf Abstract}
\end{center}

We investigate the constraints of crossing symmetry on CFT correlation functions. 
Four point conformal blocks are naturally viewed as functions on the upper-half plane, on which crossing symmetry acts by $PSL(2,\ZZ)$ modular transformations.
This allows us to construct a unique, crossing symmetric function out of a given conformal block by averaging over $PSL(2,\ZZ)$.  In some two dimensional CFTs the correlation functions are precisely equal to the modular average of the contributions of a finite number of light states.  
For example, in the two dimensional Ising and tri-critical Ising model CFTs, the correlation functions of identical operators are equal to the $PSL(2,\ZZ)$ average of the Virasoro vacuum block; this determines the 3 point function coefficients uniquely in terms of the central charge.
The sum over $PSL(2,\ZZ)$ in CFT$_2$ has a natural AdS$_3$ interpretation as a sum over semi-classical saddle points, which describe particles propagating along rational tangles in the bulk.
We demonstrate this explicitly for the correlation function of certain heavy operators, where we compute holographically the semi-classical conformal block with a heavy internal operator.

\newpage

\tableofcontents

\section{Introduction}

The conformal bootstrap is a powerful tool to constrain the spectrum and dynamics of strongly coupled field theories.  It is especially powerful in two dimensions \cite{Belavin:1984vu}, where it has led to an exact classification of the rational models \cite{Cappelli:1986hf}.  Bootstrap techniques have recently been used to place strong constraints on higher dimensional conformal field theories as well (see e.g. \cite{ElShowk:2012ht,Rychkov:2016iqz} and citations therein). 
The goal of this paper is to describe a somewhat different implementation of the conformal bootstrap program which is inspired by the modular properties of conformal blocks.  Most of our explicit computations are in two dimensions, although we expect the general strategy to apply in higher dimensions as well. 
 
We begin with the usual starting point of the conformal bootstrap: the expansion of a CFT correlation function as a sum over intermediate states.  For example, the four point function of a scalar operator $\op$ can be written as a sum over intermediate operators $\phi$ as\footnote{In this equation we are using conventions where the $|x|^{\Delta_\phi}$ appears explicitly in front of the conformal block, in order to emphasize that low dimension operators will dominate the sum when $x\to 0$.  Later we will absorb this factor $|x|^{\Delta_\phi}$ into the definition of the conformal block, as is standard in much of the literature.}
 \be\label{blocky}
\langle\op(z_1) \op(z_2)\op(z_3)\op(z_4)\rangle = G(\{z_a\}) \left(\sum_\phi C_{\op \op \phi}^2 |x|^{\Delta_\phi} {\F}_{\phi}(x, {\bar x})\right)~.
 \ee
 Here $G(\{z_a\})$ is a known function of the operator locations $z_a$, which will not be important here, and $x$ is the conformally invariant cross ratio:
 \be
 \frac{z_{12}^2z_{34}^2}{z_{13}^2z_{24}^2}=x\bar{x}\,,\quad \frac{z_{14}^2z_{23}^2}{z_{13}^2z_{24}^2}=(1-x)(1-\bar{x})~.
 \ee
The sum in \rref{blocky} is over all primary operators $\phi$, with dimensions $\Delta_\phi$, and is weighted by the square of the three point coefficient $C_{\op \op \phi}$. The conformal block  ${\F}_\phi(x, {\bar x})$ encodes the contribution of the entire family of conformal descendants of $\phi$, and is a function only of the dimension and spin of $\phi$ and $\op$.
  Equation \rref{blocky} is a general formula, but many simplifications occur in two dimensions.  In this case ${\F}_\phi(x,{\bar x})$ is the product of a left- and a right-moving block. Moreover, in $D=2$ the states can be organized into representations of Virasoro symmetry, so ${\F}_\phi(x, {\bar x})$ can be taken to be the full Virasoro block rather than just a global conformal block.

The basic observation is that the four point function
$\langle\op(z_1) \op(z_2)\op(z_3)\op(z_4)\rangle$ 
must be invariant under crossing symmetry, i.e. invariant under permutations of the operators $O(z_a)$.  The expansion \rref{blocky} is not manifestly invariant under crossing symmetry -- the conformal blocks transform in a highly non-trivial way -- so this is a strong constraint on the operator dimensions and three point coefficients.
In the standard implementation of the conformal bootstrap, one attempts to solve this constraint directly.

The problem is that the constraints of crossing are difficult to write down explicitly.  For example, in general the dimensions $\Delta_\phi$ of the intermediate states are not known, so one must solve for these dimensions at the same time that one is solving for the three point coefficients.  However, if we are only interested in the limit $x\to 0$, the sum is dominated by the identity operator, so that\footnote{In $D>2$ the identity block is trivial, but in $D=2$ we can (and will) use Virasoro blocks where ${\F}_{\id}(x, {\bar x})$ is non-trivial.} 
\be
\langle\op(z_1) \op(z_2)\op(z_3)\op(z_4)\rangle 
\approx
G(\{z_i\}) {\F}_{\id}(x, {\bar x}) 
+ \dots
\ee
More generally, when $x$ is small we can approximate the four point function by
\be\label{light}
\langle\op(z_1) \op(z_2)\op(z_3)\op(z_4)\rangle 
\approx
G(\{z_i\}) {\F}_{\text{light}}(x, {\bar x})
\ee
where
\be{\F}_{\text{light}}(x, {\bar x})\equiv \sum_{\Delta_\phi < \Delta_{light}} C_{\op \op \phi}^2 |x|^{\Delta_\phi} {\F}_{\phi}(x, {\bar x})
\ee
is the contribution from the ``light" operators, i.e. the operators with dimension less than some value $\Delta_{light}$:
This approximation has the advantage that it requires only CFT data involving light operators.
As we increase $\Delta_{light}$ the approximation \rref{light} becomes more accurate, but requires more detailed information about the CFT.

Our strategy is motivated by the following question: 
Given only the contribution ${\F}_{light}(x, {\bar x})$ from a set of light states, can we construct a consistent ``candidate" correlation function 
\be
\langle\op(z_1) \op(z_2)\op(z_3)\op(z_4)\rangle_{\text{candidate}}
\ee 
which has all of the desired properties of a true CFT four point function?  
In particular, we will
seek a candidate correlation function that
\begin{itemize}
	\item matches the  $x\to 0$ behaviour of the light operators in \rref{light},
	\item is crossing symmetric, and
	\item is a single valued function of the cross ratio $x$.
\end{itemize}
The first property is easy to satisfy.  We can just take our candidate partition function to be the truncated sum \rref{light}, which includes only light operators.  The second property is, at least naively, just as straightforward: one could simply sum the result over all possible permutations of the external operators.  This has the effect of summing over channels in which the intermediate light states could propagate.  The real problem is the third condition.  The conformal blocks are not single valued functions of $x$; they have branch cuts with non-trivial monodromy structure around $x=1$, which is the radius of convergence of the OPE expansion \rref{blocky}.

We propose to resolve this problem by exploiting the modular structure of conformal blocks.  In particular, we will use the fact that conformal blocks are naturally viewed as functions not of cross ratio, but rather as functions of a modular parameter $\tau$ which lives on the upper half plane $\HH_+$.  This observation has appeared in the literature before (see e.g. \cite{Zamolodchikov1987conformal}), and will be reviewed in detail in the next section.  The upper half plane $\HH_+$ is the universal cover of $x$-space, so the conformal blocks are single valued functions of $\tau$.  In this language, crossing symmetry is simple to state: regarded as a function of $\tau$, the four point function \rref{blocky} must be invariant under the modular transformations
\begin{equation}
\tau \mapsto \gamma \tau \equiv \frac{a \tau + b}{c \tau +d},~~~~~~{\rm for~all~} \gamma = \begin{pmatrix}a&b\\c&d\end{pmatrix}\in SL(2,\ZZ)~.
\end{equation}
Our question can therefore be rephrased as follows: Given a light contribution ${\F}_{\text{light}}(\tau, {\bar \tau})$, how do we turn it into a modular invariant function of $\tau$?

Our proposal is that candidate correlation functions should be constructed by averaging the light contribution ${\F}_{\text{light}}(\tau, {\bar \tau})$ over the modular group $PSL(2,\ZZ)$:
\be\label{farey}
\langle\op(z_1) \op(z_2)\op(z_3)\op(z_4)\rangle_{\text{candidate}} =G(\{z_i\}) \frac{1}{N}\sum_{\gamma\in PSL(2,\ZZ)} {\F}_{\text{light}}(\gamma\tau, \gamma {\bar \tau}),
\ee
where $N$ is a normalization constant.  This average, provided it converges, satisfies all of our criteria.  It can be viewed as an improved sum of the light contribution ${\F}_{light}(x, {\bar x})$ over all possible channels, including those obtained by non-trivial monodromies of the cross-ratio.   In the mathematics literature, averages over $PSL(2,\ZZ)$ are known as Poincar\'e series (see e.g. \cite{rademacher1939fourier, apostol2012modular}).  In the physics literature they are often referred to as Farey tail sums, and have appeared primarily in the context of three dimensional gravity (see e.g. \cite{Dijkgraaf:2000fq, Kraus:2006wn, Maloney:2007ud, Duncan:2009sq, Castro:2011xb, Keller:2014xba}).
Unfortunately, in many cases sums of the form \rref{farey} will diverge, and must be regulated; regularizations of sums of this type were considered in \cite{Dijkgraaf:2000fq, Manschot:2007ha, Maloney:2007ud, Keller:2014xba}.
In this work we will focus on this sum primarily in the context of minimal models, where the convergence is manifest.

One can view the proposal \rref{farey} as a construction of an approximate four point function which has the advantage that it depends only on the light data of the theory, i.e. on the dimensions and three point coefficients of operators with $\Delta_\phi < \Delta$.  The two dimensional case is particularly interesting, because in this case the Virasoro vacuum block itself is non-trivial.  So one can take ${\F}_{\text{light}}(x, {\bar x}) = {\F}_{\id}(x, \bar{x})$, including in the sum only the contribution of the vacuum block. This gives candidate four point functions which are determined uniquely in terms of the central charge.  Even in higher dimensions, one can imagine including only the contributions of the stress tensor or of other universal light operators as a seed contribution from which to construct the candidate correlation functions\footnote{If one takes only the vacuum block as the seed contribution, the sum contains three terms, being the product of two-point functions in S,T and U channels. This gives the disconnected piece of the correlation function, the generalised free field result.}.

The utility of this approach becomes clear when we imagine taking the candidate four point function \rref{farey} and re-expanding around $x\to0$ as in \rref{blocky}.  In this case we can ask the following: does the candidate four point function reproduce the contribution of heavy states as well?  In particular, by expanding around $x\to0$ one can attempt to extract from our candidate four point function the dimensions and three point coefficients of other operators in the theory. In general there is no guarantee that the resulting coefficients $C_{\op \op \phi}$ extracted in this way would be real.  In this case one would discover that additional heavy operators need to be added at a particular dimension. This would provide a novel implementation of the conformal bootstrap strategy.

On the other hand, one might hope that in some cases the candidate four point function constructed in \rref{farey} might be exactly correct. 
This would be a truly miraculous occurrence, since by re-expanding 
around $x\to 0$ and using \rref{blocky} one could then read off the dimensions and three-point coefficients of all operators of the theory.  We will see that, for rational CFTs in two dimensions, miracles do indeed occur.  For example, in the 2D Ising model CFT we will see that \emph{all} of the correlation functions of the theory are given by modular sums \rref{farey}, where we include only the Virasoro vacuum block in ${\F}_{\text{light}}(x, {\bar x})= {\F}_\id (x, {\bar x})$.\footnote{This property was previously observed for the Ising model partition function in \cite{Castro:2011zq}; our result is an extension of this to correlation functions.}  In analogy with \cite{Witten:2007kt}, we will call a CFT correlation function which has the property that it is equal to the modular average of the vacuum block an ``extremal correlator."  If a CFT correlator is extremal, then all of the three-point coefficients are determined in terms of the central charge.  We will show explicitly that the Ising model correlators are extremal, and present numerical evidence that other minimal model correlators are extremal as well.

The sum over $PSL(2,\ZZ)$ described above has, at least in some cases, a natural AdS/CFT interpretation as a sum over semi-classical  bulk saddles.  One saddle point contribution to a CFT four point function is described by a pair of bulk worldlines which connect the two pairs of boundary points. 
 For two dimensional CFTs (three dimensional bulk) these worldlines can be topologically non-trivial.  We will see that the sum over $PSL(2,\ZZ)$ corresponds precisely to the sum over particle worldlines which map out ``rational tangles" in the bulk.  
When the boundary operators are heavy the worldlines will back-react on the geometry, so the sum over bulk saddles is difficult to compute precisely.
 We will therefore focus on a particular computation -- that of external operators of dimension $\Delta=\frac{c}{16}$ -- where the computation can be performed explicitly.  In this case we will show that the holographic computation of the correlation function takes precisely the form of a sum over rational tangles.  
 This is holographic evidence that, at least in some cases, our candidate correlation functions constructed by a modular average are correct.

We note that our computation of the correlation function of $\Delta=\frac{c}{16}$ operators includes new results on the AdS gravity interpretation of conformal blocks.  Our semiclassical, first-quantised description of particles in AdS$_3$ will naturally compute Virasoro conformal blocks in the boundary CFT.  In particular, we will compute from gravity an exact (in dimensions, leading order in $c$) semiclassical block, where all external and internal operators have dimensions of order $c$. 

In section \ref{sec:crossingmodular} we will review the relationship between modular transformations and crossing symmetry.  In section \ref{sec:poincaresum} we will describe in detail our proposal for the candidate correlation function as a modular average.  In section \ref{sec:minimalmodel} we will demonstrate that the candidate correlation functions are exactly correct in certain two dimensional rational CFTs, but that they fail to be exact in other cases.  In section \ref{sec:semiclassical} we describe the interpretation of the Farey tail approximation in AdS/CFT, where the sum over $PSL(2,\ZZ)$ can be regarded as a sum over saddle point contributions to a semi-classical correlation function.
We also give gravitational interpretation of the conformal block when $\Delta = \frac{c}{16}$.  Much of this section can be read independently from the rest of the paper.
We emphasize that, although the general modular structure applies in all dimensions $D\ge 2$, in this paper we will describe specific computations of the Farey tail sum only in $D=2$.

While this paper was in preparation, we learned that related results will be discussed in \cite{CGL,Balasubramanian:2017fan}.  We thank these authors for their correspondence.

\section{Crossing symmetry as modular invariance}
\label{sec:crossingmodular}
In this section we will describe the relationship between crossing symmetry and modular invariance.  This section is largely a review, although in much of the literature this relationship is not discussed explicitly.

\subsection{Four-point functions and crossing symmetry}

We are interested in the 4-point correlation function
\begin{equation}
\langle\op_1(z_1)\op_2(z_2)\op_3(z_3)\op_4(z_4)\rangle
\end{equation}
of a $D$ dimensional conformal field theory in Euclidean signature. This is a function of 4 points $\{z_a\}$ which transforms covariantly under conformal transformations. We can use these conformal transformations to send $z_4$ to infinity, $z_1$ to the origin, $z_3$ to $(1,0,\ldots,0)$ and $z_2$ to a point in the $x_1-x_2$ plane. Using complex coordinates, we will denote the resulting position of $z_2$ by $x\in\CC-\{0,1\}$.\footnote{In $D>2$ we have the additional freedom to choose $\Im{x}\geq 0$, but we will not insist on this here.} The parameter $x$ is the cross-ratio, which can be written in terms of the original four points as
\begin{equation}
u=\frac{z_{12}^2z_{34}^2}{z_{13}^2z_{24}^2}=x\bar{x}\,,\quad v=\frac{z_{14}^2z_{23}^2}{z_{13}^2z_{24}^2}=(1-x)(1-\bar{x})
\end{equation}
where $z_{ab}=z_a-z_b$. In two dimensions we simply have 
\be
x=\frac{z_{12}z_{34}}{z_{13}z_{24}}.
\ee
The configuration space of our four distinct marked points $\{z_a\}$, modulo conformal transformations, is the twice-punctured plane (or thrice punctured sphere): $x\in\CC-\{0,1\}$. In two dimensions this space is usually denoted ${\cal M}_{0,4}$, the moduli space of four points on $S^2$.  We will simply refer to this space as $x$-space, or cross-ratio space.

Our four point function can be written as:
\begin{equation}\label{bilbo}
\langle\op_1(z_1)\op_2(z_2)\op_3(z_3)\op_4(z_4)\rangle = G_0\left(\{z_a\};\{\Delta_a,s_a\}\right) G_{1234}(x, {\bar x}).
\end{equation}
Here $G_0$ is a function chosen once and for all, containing only kinematic data, which transforms like a correlation function under conformal transformations.  It is convention dependent and depends only on the dimensions $\Delta_a$ and spins $s_a$ of the operators $\op_a$.  For operators with spin in $D>2$, Equation \rref{bilbo} should include a sum over multiple terms, one for each different tensor structure that can appear in the correlator.  We will focus on scalar operators where this is not an issue.
For $D=2$ we will choose $G_0$ to contain the correct branch structure encoding the statistics of anyonic operators, so $G_{1234}(x,{\bar x})$ is single-valued. For scalar operators of dimension $\Delta_i$ we will choose 
\be
G_0\left(\{z_a\};\{\Delta_a\}\right) = 
\prod_{a<b}|z_{ab}|^{\Delta/3-\Delta_a-\Delta_b} 
\ee
where $\Delta=\sum_a\Delta_a$. 
This convention has the advantage that $G_0$ treats the operators democratically, in the sense that it is invariant under permutation of the $O_a(z_a)$.

Crossing symmetry is the invariance of the four point function under permutations of the operators $\op_a(z_a)$.
The cross ratio transforms under this permutation, and as a result the functions $G_{abcd}(x, {\bar x})$ are related by
\bea\label{crossing}
G_{1234}(x, {\bar x}) &=& G_{1243}\left(\frac{x}{x-1},\frac{\bar x}{{\bar x}-1}\right) = G_{4231}\left(\frac{1}{x},\frac{1}{\bar x}\right)\nonumber 
= G_{4213}\left(\frac{x-1}{x},\frac{{\bar x}-1}{\bar x}\right) 
\\
&=& G_{3241}\left(\frac{1}{1-x},\frac{1}{1-{\bar x}}\right) = G_{3214}\left(1-x,1-{\bar x}\right).
\eea
The permutation of the indices, and the action on $x$, come from the application of six M\"obius maps that permute $z_1,z_3,z_4$.  These six permutations form the anharmonic group, which is isomorphic to the symmetric group $S_3$. The remaining permutations which interchange $z_2$ with one of $z_1,z_3,z_4$ give\footnote{One may notice that these are the same as the symmetries of the Riemann tensor. This is no accident: for exactly marginal operators, an integrated four-point function gives the curvature of moduli space.} 
\be
G_{1234}(x, {\bar x})=G_{2143}(x, {\bar x})=G_{3412}(x, {\bar x})=G_{4321}(x, {\bar x})~.
\ee

\subsection{The upper half plane as the universal cover}

Conformal blocks are not single-valued functions of the cross-ratio $x$, so it will be convenient to pass to the universal cover of cross-ratio space.  In doing so, we wish to keep the local analytic structure intact.  The essential point is that cross-ratio space, viewed as the thrice-punctured sphere, is a Riemann surface.  So, as with almost all Riemann surfaces, the universal cover is the upper half plane $\HH_+$. 

Concretely, we can take the cross ratio $x$ to be the image of a point $\tau\in \HH_+$ under the modular $\lambda$ function:
\begin{equation}
x = \lambda(\tau) =\left(\frac{\sqrt{2}\,\eta(\tau/2)\eta^2(2\tau)}{\eta^3(\tau)}\right)^8 = \left(\frac{\theta_2(\tau)}{\theta_3(\tau)}\right)^4~.
\end{equation}
The limit $x\to 0$, where the identity block dominates, is given by $\tau\to i \infty$ on $\HH_+$.  
Locally, we can write the inverse of our map as 
\begin{equation}\label{saruman}
\tau(x) = i\frac{K(1-x)}{K(x)}
\end{equation}
where $K(x)$ is the elliptic integral
\begin{equation}
K(x)=\frac{1}{2}\int_0^1 \frac{dt}{\sqrt{t(1-t)(1-x t)}}=\frac{\pi}{2}.\; {}_2F_1\left(\frac{1}{2},\frac{1}{2};1;x\right). 
\end{equation}
Of course, the inverse \rref{saruman} is not unique; it describes the infinite set of pre-images $\tau(x)$ on $\HH_+$.  This is reflected by the fact that $\tau(x)$ is a not a single valued function of $x$.

An advantage of this perspective is that cross-ratio space can now be viewed as a quotient of the upper half plane.
In particular, the modular $\lambda$ function is invariant under the action of the congruence group $\Gamma(2)$:
\be
\lambda(\tau) = \lambda(\gamma\tau),~~~~~{\rm for~all~}\gamma\in \Gamma(2)~.
\ee
Here $\Gamma(2)$ is the index 6 normal subgroup of the modular group $PSL(2,\ZZ)$, which is generated by the two M\"obius maps
\begin{equation}
T^2:\tau\mapsto\tau+2,\quad ST^2S:\tau\mapsto \frac{\tau}{-2\tau+1}~.
\end{equation}
\begin{figure}
	\centering
	\begin{subfigure}[b]{.45\textwidth}
	\includegraphics[width=.9\textwidth]{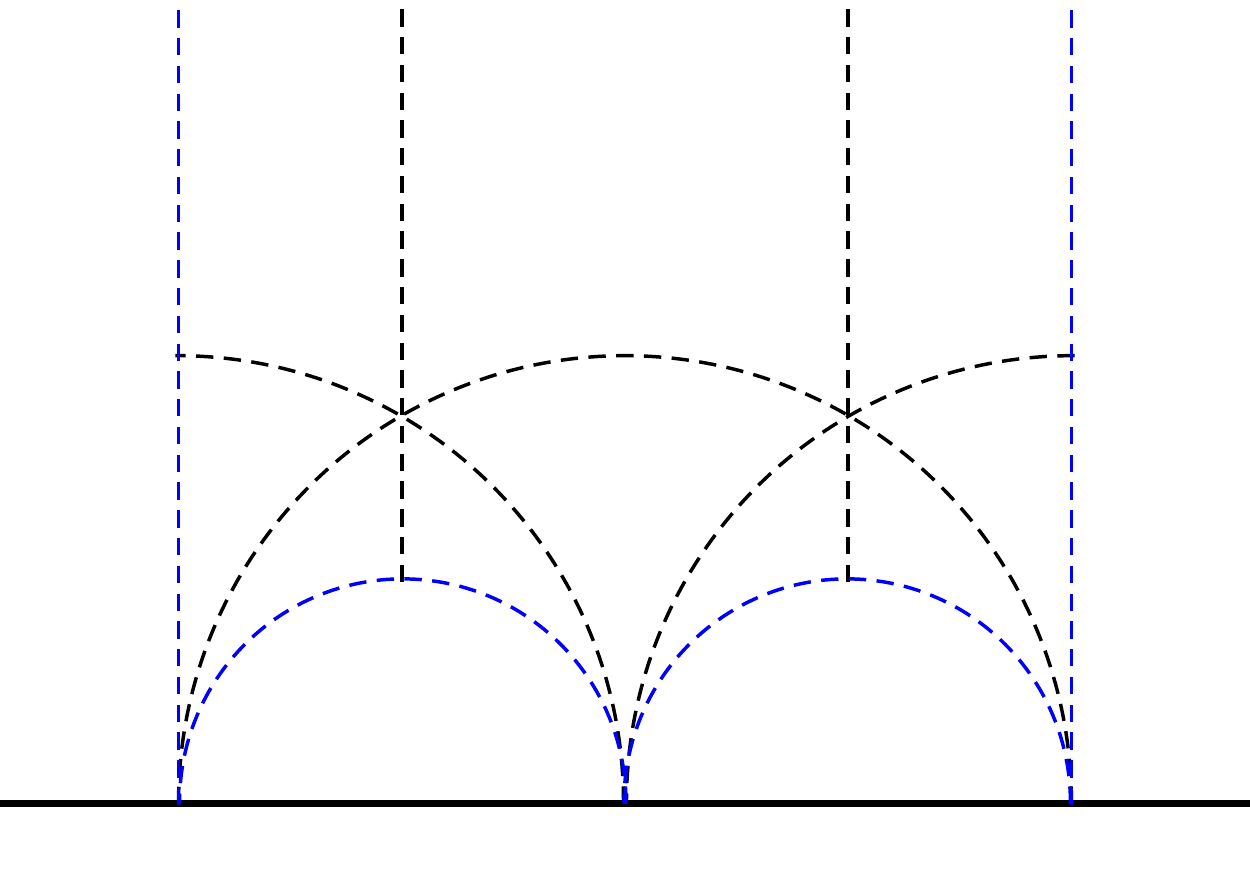}
	\caption{$\tau$-plane}
	\end{subfigure}
	\begin{subfigure}[b]{.45\textwidth}
	\includegraphics[width=.9\textwidth]{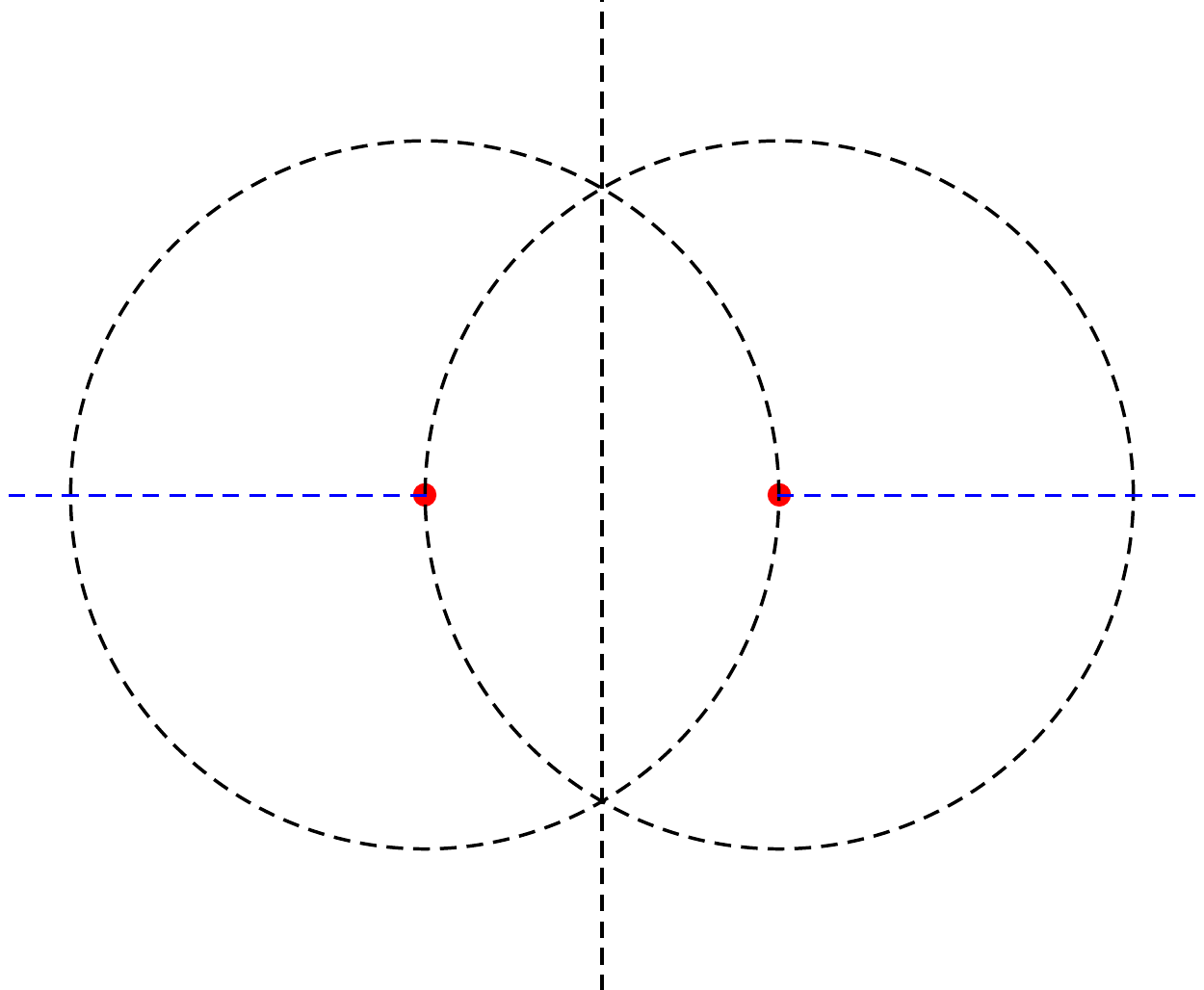}
	\caption{$x$-plane}
	\end{subfigure}
	\caption{Fundamental domain for $\Gamma(2)$ in the upper half $\tau$-plane, bounded by the blue dashed curves. $T^2$ identifies the left and right vertical lines, and $ST^2S$ identifies the semicircles. The three cusps at $\tau=i\infty$, $0$ and $\pm1$ correspond to $x=0,1,\infty$. The black dashed lines show how this domain breaks up further into six fundamental domains for the modular group $\Gamma$ (four of which are split in two across the blue lines). These six domains correspond to the images in the cross-ratio $x$-plane under the anharmonic group, shown in the right figure, where the marked points are at $x=0,1$. \label{fig:Gamma2Domain}}
\end{figure}
This means that we can view cross-ratio space as the quotient $\HH_+ / \Gamma(2)$.  
The group $\Gamma(2)$ is identified with the fundamental group of cross-ratio space (see \cref{fig:Gamma2Domain}), $\CC-\{0,1\}$, which is the free group on two elements.

We conclude that we can write the functions $G_{abcd}(x)$ of cross-ratio as functions of $G_{abcd}(\tau)$ on $\HH_+$ which are invariant under $\Gamma(2)$.

\subsection{Crossing symmetry as modular invariance}

We can now ask how the remaining modular transformations $\gamma \in PSL(2,\ZZ)$ act on the cross ratio $x=\lambda(\tau)$.  
The important point is that the generators of $PSL(2,\ZZ)$ act as crossing transformations:
\be\label{modularaction}
T\cdot x=\frac{x}{x-1},~~~~~S\cdot x = 1-x.
\ee
Indeed, the quotient $PSL(2,\ZZ)/\Gamma(2) = S_3$ is precisely the anharmonic group described above, which acts as the six nontrivial M\"obius maps on $x$ given in equation \rref{crossing}.  So crossing symmetry implies that the $G_{abcd}(\tau)$ collectively transform into one another under modular transformations.

Before considering the general case, let us first consider the special case when the four external operators $\op_a$ are identical.  
In this case, swapping identical points will leave the correlator invariant.  So $G_{abcd}(x) \equiv G(x)$ is invariant under the anharmonic group.  Thus, as a function on the upper half plane, $G(\tau)$ must be invariant under the full modular group $PSL(2,\ZZ)$. 
Any such function can be written as a function of the $j$-invariant
\begin{equation}
j(\tau) = \frac{256 (1-x(1-x))^3}{x^2(1-x)^2}
\end{equation}
which assigns a unique complex number to each point on the fundamental domain $\HH_+/SL(2,\ZZ)$.  In cross-ratio space, this fundamental domain is $\{x:|x-1|<1,\Re(x)<1/2\}$.\footnote{
In the special case where $G(\tau)$ is a meromorphic function of $\tau$, $G(\tau)$ will be a rational function of $j(\tau)$ which is uniquely determined by its poles and zeros.  This can be used to efficiently compute the correlation function of chiral operators in two dimensional CFTs, as in \cite{Headrick:2015gba}.}

When some, but not all, of the operators $O_a(z_a)$ are identical, the function $G_{abcd}(\tau)$ will be invariant under a subgroup $\Gamma\subseteq PSL(2,\ZZ)$ which contains $\Gamma(2)$.  For three identical operators it is invariant under $PSL(2,\ZZ)$.
With two identical operators, or two pairs of identical operators, it is invariant under an index 3 subgroup of $PSL(2,\ZZ)$ (the congruence subgroup $\Gamma_1(2)$), which itself contains $\Gamma(2)$ as an index 2 subgroup.

It is useful to reformulate this slightly, by regarding the  $G_{abcd}(\tau)$ collectively as the components of a vector-valued modular function $\vec{G}(\tau)$.  This means that the components of $\vec{G}(\tau)$ will in general transform into one another under a modular transformation.  This has the effect of restricting the domain to the fundamental region of $\HH_+/SL(2,\ZZ)$, at the expense of introducing multiple functions that map into one another under modular transformations. In particular, we have 
\begin{equation}\label{eq:ModularG}
\vec{G}(\gamma\tau) = \sigma(\gamma) \vec{G}(\tau)
\end{equation}
where $\sigma$ is a representation of the permutation group.  The representation is six dimensional in the general case, three dimensional when two operators or two pairs are identical, or one dimensional when three or all four operators are identical.

It is useful to write this all out explicitly.
Expressed in terms of $\tau$, crossing symmetry is
\be
G_{abcd}(\tau)=G_{bacd}(\tau+1)=G_{adcb}(-1/\tau)
\ee
along with
\be
G_{abcd}(\tau) = G_{badc}(\tau) = G_{dcba}(\tau) = G_{cdab}(\tau). 
\ee
Arranging the independent functions in a six-dimensional vector
\begin{equation}\label{vecG}
\vec{G}(\tau) = \left(G_{1234}(\tau),G_{2134}(\tau),G_{4132}(\tau),G_{1432}(\tau),G_{2431}(\tau),G_{4231}(\tau)\right)^t
\end{equation}
the crossing relations can be written as
\begin{equation}\label{sigmais}
\renewcommand*{\arraystretch}{.9}
\setlength{\arraycolsep}{2.6pt}
\vec{G}(\tau+1) =
\begin{pmatrix}
0 & 1 & 0 & 0 & 0 & 0\\
1 & 0 & 0 & 0 & 0 & 0\\
0 & 0 & 0 & 1 & 0 & 0\\
0 & 0 & 1 & 0 & 0 & 0\\
0 & 0 & 0 & 0 & 0 & 1\\
0 & 0 & 0 & 0 & 1 & 0
\end{pmatrix}\vec{G}(\tau) ;\; \vec{G}(-1/\tau) =
\begin{pmatrix}
0 & 0 & 0 & 1 & 0 & 0\\
0 & 0 & 0 & 0 & 1 & 0\\
0 & 0 & 0 & 0 & 0 & 1\\
1 & 0 & 0 & 0 & 0 & 0\\
0 & 1 & 0 & 0 & 0 & 0\\
0 & 0 & 1 & 0 & 0 & 0
\end{pmatrix}\vec{G}(\tau)~.
\end{equation}
This is a reducible representation of the anharmonic group $S_3$.  It is the sum of the trivial representation, the one-dimensional sign representation and two copies of the `standard' two-dimensional representation. One basis for this decomposition is
\begin{align*}
G^{\mathrm{triv}}(\tau)&=\frac{1}{\sqrt{6}}\left(G_{1234}(\tau)+G_{2134}(\tau)+G_{4132}(\tau)+G_{1432}(\tau)+G_{2431}(\tau)+G_{4231}(\tau)\right)\\
G^{\mathrm{sign}}(\tau)&=\frac{1}{\sqrt{6}}\left(G_{1234}(\tau)-G_{2134}(\tau)+G_{4132}(\tau)-G_{1432}(\tau)+G_{2431}(\tau)-G_{4231}(\tau)\right)\\
\vec{G}^{\mathrm{std}}_1(\tau)&=\frac{1}{\sqrt{3}}\begin{pmatrix}
G_{1234}(\tau)+\omega G_{4132}(\tau)+\omega^2G_{2431}(\tau)\\[.8 em]
\omega^2 G_{2134}(\tau)+G_{1432}(\tau)+\omega G_{4231}(\tau)
\end{pmatrix}\\
\vec{G}^{\mathrm{std}}_2(\tau)&=\frac{1}{\sqrt{3}}\begin{pmatrix}
\omega G_{2134}(\tau)+G_{1432}(\tau)+\omega^2 G_{4231}(\tau)\\[.8 em]
G_{1234}(\tau)+\omega^2 G_{4132}(\tau)+\omega G_{2431}(\tau)
\end{pmatrix}
\end{align*}
where $\omega=e^{2\pi i/3}$. Under the modular group, the trivial representation is invariant, the sign representation picks up a $(-1)$ from the action of $S$ or $T$, and the standard representation in the chosen basis is
\begin{equation}
\rho^\mathrm{std}(T) = \begin{pmatrix}
0 & \omega \\
\omega^2 & 0 \\
\end{pmatrix};\quad \rho^\mathrm{std}(S) = \begin{pmatrix}
0 & 1 \\
1 & 0 \\
\end{pmatrix}.
\end{equation}
A correlation function be described as the collection of four vector valued modular functions for $PSL(2,\ZZ)$, transforming in the above representations.

When some of the operators are identical, we may not need all these representations.  If all four operators are identical, three representations identically vanish, and only the trivial representation remains. If $\op_1\equiv\op_2$ and $\op_3\equiv\op_4$, the sign representation vanishes and the two copies of the standard representation are proportional, leaving the trivial representation and one two-dimensional representation.

The description of crossing symmetry as modular transformations is not particularly useful when discussing the correlation functions themselves; we have just replaced the anharmonic group of crossing symmetries with the infinite dimensional modular group $PSL(2,\ZZ)$.  For the four point functions themselves, this extra structure is not necessary.  The advantage of the present approach is that -- because $\HH_+$ is the universal cover of cross-ratio space -- the conformal blocks are single valued functions of $\tau$, even though they are multiply valued functions of $x$.

\section{A Poincar\'e series for correlation functions}
\label{sec:poincaresum}

We can now describe our construction of a candidate four point function as a sum over the modular group $PSL(2,\ZZ)$.
The expansion of the four point function \rref{bilbo}
as a sum over intermediate states takes the form
\begin{equation}\label{frodo}
G_{abcd}(x, {\bar x}) = \sum_p C^p_{ab}C^p_{cd} \F_p^{ab,cd}(x, {\bar x}),
\end{equation}
where ${\F}_p^{ab,cd}(x, {\bar x})$ is the conformal block associated with the primary operator $\op_p$.  
We are still working in general $D\ge 2$, although we will later specialize to the case $D=2$ where ${\F}_p^{ab,cd}(x, {\bar x})$ will be the product of left- and right-moving Virasoro blocks.
We have absorbed into ${\F}_p^{ab,cd}(x, {\bar x})$ the usual factors of $x$, so that
\be
{\F}_p^{ab,cd}(x, {\bar x}) \sim |x|^{\Delta_{p}-\Delta/3}
+\cdots
\ee
as $x\to 0$. In the case of a four point function of identical scalar operators, the conformal block defined here is  $|x|^{\Delta_p}$ times the conformal block in \cref{blocky}.

If we wish to approximate our four point function at $x\to0$, it is sufficient to include only the contributions to $G_{abcd}$ from low-lying operators, i.e. to take
\be
G_{abcd}(x,{\bar x}) = F^{\text{light}}_{ab,cd}(x,{\bar x}) + \cdots
\ee
where
\be
F^{\text{light}}_{ab,cd}(x,{\bar x}) = \sum_{\Delta_p\le \Delta_{light}}  C^p_{ab}C^p_{cd} \F_p^{ab,cd}(x, {\bar x})
\ee
includes only contributions from operators below some dimension $\Delta$.  For $D=2$, even the Virasoro vacuum block contribution is non-trivial. 

In the notation of the previous section, where we assemble the six $G_{abcd}$ into a vector according to \rref{vecG}, we can write this more succinctly as 
\be\label{sam}
{\vec G}(x, {\bar x}) = {\vec F}^{\text{light}}(x, {\bar x}) + \cdots
\ee 

In our expansion \rref{frodo}  
the four point function $G_{abcd}(x, {\bar x})$ is a single valued function of $x$. The individual conformal blocks, however, are not.  
They have non-trivial monodromies as one moves around in cross-ratio space.  Of course, this 
branch structure will disappear when we perform the sum over $p$ in \rref{frodo} to obtain the single valued $G_{abcd}$. 

This branch structure means that the conformal blocks should be regarded as function of the covering coordinate $\tau$ rather than $x$.  
From \rref{modularaction} we see that the monodromy around $x=0$ is generated by the shift $T^2:\tau\to\tau+2$ and that the monodromy around $x=1$ is generated by $ST^2S$.  This allows us to completely unwrap the branch structure and view $\F_p^{ab,cd}(\tau, {\bar \tau})$ as a single valued function of $\tau$. 
 
Our approximate four point function \rref{sam} should therefore really be written as an equation on $\HH_+$, as 
 \be\label{samwise}
 \vec{G}(\tau, {\bar \tau}) = {\vec F}^{\text{light}}(\tau, {\bar \tau}) + \cdots
 \ee
 Our goal is then to ask how this can be completed to a crossing symmetric four point function.  In particular, we will fix the $\cdots$ terms in \rref{samwise} by demanding that the four point function obeys
 \be\label{cat}
\vec{G}(\gamma\tau, \gamma{\bar \tau}) =
 \sigma(\gamma)\cdot
 \vec{G}(\tau, {\bar \tau})
 \ee
 where $\sigma(\gamma)$ is the six dimensional representation of the modular group defined in equation \rref{sigmais}.
Our ansatz is that we simply average over the modular group, by setting
 \be\label{candidate}
 \vec{G}_{\text{candidate}}(\tau, {\bar \tau}) = \frac{1}{N}\sum_{\gamma\in PSL(2,\ZZ)} \sigma^{-1}(\gamma)\cdot{\vec F}^{\text{light}}(\gamma \tau, \gamma {\bar \tau}).
 \ee
 Here $N$ is a normalization constant, which is fixed by demanding that the limit $x\to 0$ ($\tau \to i \infty$) matches with the light limit \rref{samwise}.  
Provided the sum converges, $\vec{G}_{\text{candidate}}$ automatically obeys \rref{samwise} and \rref{cat}.
 The ansatz \rref{candidate} should be viewed as a precise version of the statement that heavy states arise through the propagation of light states in a dual channel.
  
 In the next section we will unpack this statement and perform specific computations with this ansatz for two dimensional minimal model CFTs.
 Before proceeding, however, we should make a few comments. Sums of this sort appear frequently in number theory, both in the holomorphic and non-holomorphic settings.  They have also been considered extensively in the context of three dimensional gravity.  One important feature is that the convergence of the sum \rref{candidate} is not guaranteed, and the regularization can be quite subtle (see e.g. \cite{Dijkgraaf:2000fq, Manschot:2007ha, Maloney:2007ud, Keller:2014xba}). 
 In some cases the sum can only be defined using zeta function regularization, and the normalization constant $N$ is formally infinite.  
In some of the explicit computations performed below, however, the sum will collapse to a finite sum in an obvious way, so convergence will not be an issue; a similar phenomenon was noted in \cite{Castro:2011zq}.  
 
\section{Minimal models}
\label{sec:minimalmodel}
We will now construct correlation functions by performing the modular average explicitly in some unitary minimal models in $D=2$. We begin by recalling a few facts on the 2D minimal models\footnote{Our discussions and conventions are based on \cite{francesco2012conformal}.}.
For a pair of coprime integers $p$ and $p'$ with $p>p'$, the minimal model $M(p,p')$ has central charge
\begin{equation}
	c=1-\frac{6(p-p')^2}{pp'}.
\end{equation}
The allowed holomorphic dimensions of Virasoro primary operators are labelled by integers $r,s$, as\begin{equation}
	h_{(r,s)}=\frac{\left(p r - p' s\right)^2-(p-p')^2}{4 p p'},\quad 1\le r < p' ~\mbox{and} ~1\le s < p\,
\end{equation}
with the redundancy $h_{(r,s)}=h_{(r+p',s+p)}=h_{(p'-r,p-s)}$. We may denote such a primary by $\phi_{(r,s)}$ in a context where only the holomorphic properties are important. The physical spectrum consists of a collection of primary operators with appropriate holomorphic and antiholomorphic dimensions (constrained by modular invariance of the torus partition function), 
for example the diagonal series, for which each scalar with allowed dimension $h=\bar{h}=h_{(r,s)}$ appears exactly once.
We shall concentrate almost exclusively on the unitary series, for which $p'=p-1$. 

The section will begin with the discussion of some useful mathematical structure of the space of conformal blocks, action of the modular group, crossing symmetric four-point functions and the construction of the modular average. This will specialise the discussion in the previous section to the case with a finite number of primary operators. Next, we will move on to examples in minimal models. At the end of the section, we will present an alternative group theoretic perspective on the modular sum, motivated by results for compact groups.

\subsection{Mathematical structure}

In a two dimensional CFT a conformal block can be written as a product of holomorphic and anti-holomorphic factors, as 
\be
\F_p^{ab,cd}(x, {\bar x}) = {\mathcal F}_p^{ab,cd}(x) {\bar{\mathcal F}}_p^{ab,cd}({\bar x}).
\ee
The holomorphic and anti-holomorphic blocks ${\mathcal F}$ and ${\bar {\mathcal F}}$ depend only on the left- and right-moving dimensions of the operators, respectively.

If we fix the external operators $abcd$, we can think of the holomorphic blocks as elements of a vector space $B$ (additionally labelled by the external operator dimensions, though we will leave this implicit) with basis ${\mathcal F}_p^{ab,cd}$ labelled by the holomorphic dimension of the exchanged operator.\footnote{In this section, we restrict our discussion to finite-dimensional spaces of conformal blocks. For work and subtleties related to extension to the infinite-dimensional spaces, see \cite{Hogervorst:2017sfd}.} The antiholomorphic blocks live in the complex conjugate vector space $\bar{B}$. The correlation function is a sum of products of holomorphic and antiholomorphic blocks, which in this language is an element of $B\otimes\bar{B}$. The coefficients in the given basis are simply the products of OPE coefficients $C^p_{ab}C^p_{cd}$ (summed over all exchanged operators with the same dimensions). We may give the correlation function by arranging these in a matrix $C$, with rows and columns labelled by holomorphic and antiholomorphic dimensions respectively. Scalar operators appear on the diagonal, and operators with spin away from the diagonal.  Explicitly we have
\begin{equation}
G(x,\bar{x}) = \sum_{h,\bar{h}} \calF_{h}(x) C^{h\,\bar{h}} \bar{\calF}_{\bar{h}}(\bar x), \quad\text{where}\quad C^{h\,\bar{h}}=\!\!\!\!\sum_{\op_p:\left\{\substack{h_p=h\\ \bar{h}_p=\bar{h}}\right\}} C^p_{ab}C^p_{cd}
\end{equation}
where the last sum runs over the OPE coefficients of all operators $\op_p$ with the given holomorphic and antiholomorphic dimensions.

Concentrating firstly on the case of identical external operators, modular transformations act linearly on $B$, turning $B$ into a representation of the modular group. On the correlation functions, living in $B\otimes\bar{B}$, the modular group acts by $\gamma:C\mapsto \gamma\cdot C\cdot \gamma^\dag$, so, in particular, crossing symmetric correlation functions obey $\gamma\cdot C\cdot \gamma^\dag = C$, for $\gamma$ in the appropriate representation of $PSL(2,\ZZ)$.

Now, so far in the discussion, $B$ could include exchange of any dimension $h$, in which case it is an infinite-dimensional space, in the worst case perhaps even the space of all holomorphic functions on the upper half plane. But if we start with minimal model central charge and dimensions, the action of the modular group only produces other dimensions $h_{r,s}$: the action is highly reducible, and the finite-dimensional subspace including only the $h_{r,s}$ in the exchange is invariant under the action. In many cases, including examples below, we can further reduce the representation so that some $h_{r,s}$ do not appear, taking $B$ to be the minimal invariant space including our `seed'. This is the way in which the exchange spectrum and fusion rules appear in our construction: we do not put these data in by hand, but rather they come out as the set of exchanged blocks generated by modular images of the seed.

With the appropriate representation in hand, our proposed solution to crossing is to start with a seed and sum over images:
\begin{equation}
\label{eq:modularsumMM}
C_{\text{candidate}}\propto \sum_{\gamma \in PSL(2,\ZZ)} \gamma\cdot C_\mathrm{seed} \cdot \gamma^\dag.
\end{equation}
In the simplest case, the seed $C_\mathrm{seed}$ contains just the contribution of the vacuum block. 
In general the Virasoro vacuum block will be invariant under some subgroup $\Gamma_{stab} \subseteq PSL(2,\ZZ)$.
In this case we need sum only over the coset
\be
\label{eq:modularsumMMM}
C_{\text{candidate}}\propto \sum_{\gamma \in PSL(2,\ZZ)/\Gamma_{stab}} \gamma\cdot C_\mathrm{seed} \cdot \gamma^\dag .
\ee
In the simple cases considered below, we will see that $\Gamma_{stab}$ is often a finite index subgroup of $PSL(2,\ZZ)$.  Thus the sum has only a finite number of terms and can be computed explicitly.

It is straightforward to generalise this discussion when the external operators are not identical: we must simply add an additional label to the matrix $C$ to identify the permutation of the external operators (or the irreducible representation of $S_3$, as in \cref{vecG}), and allow the modular group to act additionally by permuting these labels. Abstractly, the space $B$ breaks up into the tensor product of a representation of $PSL(2,\ZZ)$ and a representation of the anharmonic group $S_3$.

%%%%%%%%%%%% 
\subsection{Examples}
In this section, we will perform the modular sum in \cref{eq:modularsumMM}  in three minimal model examples.
In these cases, the seed $C_\mathrm{seed}$ will be taken to be the contribution of the Virasoro vacuum block.

We will study the $p=4,5$ and $12$ diagonal minimal models.
The first two, $M(4,3)$ and $M(5,4)$, are the Ising and tricritical Ising models.  The third one, $M(12,11)$, is a coset model which will be described below. The Ising and tricritical Ising models are the simplest examples in the diagonal series where the modular average of the vacuum block correctly reproduces all of the identical operator four point functions, allowing us to uniquely determine the three point coefficients.
The diagonal $M(12,11)$ is included as an example where the modular average of the vacuum block alone fails to reproduce the three point coefficients.

\subsubsection{Ising model}
\label{sec:isingaverage}
The Ising model is the $p=4$ unitary minimal model, with central charge
$c=\frac{1}{2}$. The spectrum includes three scalar primary operators, the identity $\id\equiv \phi_{(1,1)}$, the spin field $\sigma\equiv \phi_{(1,2)}$ and the energy density $\epsilon \equiv \phi_{(2,1)}$ with dimensions
\begin{equation}
h_{(1,1)}=0,\quad
h_{(1,2)}=\frac{1}{16},\quad
h_{(2,1)}=\frac{1}{2}
\end{equation} and $\bar{h}=h$. 
The fusion rules are
\begin{equation}
\sigma \times \sigma = \id+\epsilon,\quad
\sigma \times \epsilon=\sigma,\quad
\epsilon\times \epsilon =\id.
\end{equation}
We will first consider the correlation functions of identical operators,
\begin{equation}
\langle \sigma (x_1)\sigma(x_2)\sigma(x_3) \sigma(x_4)\rangle
,\quad
\langle \epsilon (x_1)\epsilon(x_2)\epsilon(x_3) \epsilon(x_4)\rangle\,.
\end{equation}
Of these two, the fusion rule $\epsilon\times \epsilon =\id$ implies that the (holomorphic times antiholomorphic) vacuum block is already modular invariant, so the sum over $PSL(2,\ZZ)$ is trivial. We will therefore focus on the $\sigma$-four-point function.

For the mixed four-point functions, we will perform a similar computation where the following three correlators
\begin{equation}
\langle\sigma(x_1)\sigma(x_2)\epsilon(x_3)\epsilon(x_4)\rangle\, ,\quad
\langle\sigma(x_1)\epsilon(x_2)\epsilon(x_3)\sigma(x_4)\rangle\, ,\quad
\langle\sigma(x_1)\epsilon(x_2)\sigma(x_3)\epsilon(x_4)\rangle
\end{equation}
are assembled into the components of a vector-valued modular function.

\paragraph{Four-point function of $\sigma$:}\mbox{}\\

For the $\sigma$ operator four point function, there are two relevant blocks:
\begin{equation}
\calF^{\sigma\sigma\sigma\sigma}_\id=\frac{1}{\sqrt{2}}\frac{\sqrt{1+\sqrt{1-x}}}{ \left({(1-x) x}\right)^{1/12}},~~~~~~\calF^{\sigma\sigma\sigma\sigma}_\epsilon=\sqrt{2}\frac{ \sqrt{1-\sqrt{1-x}}}{\left({(1-x) x}\right)^{1/12}}.
\end{equation}
In the basis $\{ \calF^{\sigma\sigma\sigma\sigma}_\id,\calF^{\sigma\sigma\sigma\sigma}_{\epsilon}\}$ the generators $S$ and $T$ are represented by
\begin{equation}\label{isingsigmarep}
	T=e^{-i\pi/12}
	\begin{pmatrix}
		1 & 0 \\
 		0 & i
	\end{pmatrix},\quad
	S=\frac{1}{\sqrt{2}}
	\begin{pmatrix}
		1 &  2 \\
 		\frac{1}{2} & -1
	\end{pmatrix}.
\end{equation}
This matrix notation\footnote{\label{footnote1}The order of the indices here is (perhaps despite appearances) natural, because the blocks are the basis vectors of the space $B$, and so transform with a transpose relative to the components. This requires care, since it is different from conventions in much of the literature.} means that $S$ acts on $\calF_j(x)$ by $\calF_i(1-x)=\sum_j S_{ji}\calF_j(x)$.
We refer the reader to the appendix for details on calculation of these blocks and matrices.

The vacuum block has a finite orbit under the action of the modular group, represented as the six matrices listed in \cref{table1}, and the vacuum block is invariant under
the index 6 subgroup $\Gamma_{stab}=\Gamma_1(4)\subseteq PSL(2,\ZZ)$.\footnote{This itself is a subgroup of the kernel of the representation \eqref{isingsigmarep}, which is $\Gamma_0(4)$, with index 24.}
\begin{table}
  \begin{center}
  \begin{equation*}
	\begin{array}{r | c}
	\gamma & \gamma\cdot C_\mathrm{seed} \cdot \gamma^\dag \\[.2cm]
		\hline \\[-.3cm]
	\id & \begin{pmatrix}1&0\\ 0&0\end{pmatrix}\\
	S & \frac{1}{8}\begin{pmatrix}4&2\\ 2&1\end{pmatrix}\\
	TS & \frac{1}{8}\begin{pmatrix}4&-2i\\ 2i&1\end{pmatrix}\\
	T^2S & \frac{1}{8}\begin{pmatrix}4&-2\\ -2&1\end{pmatrix}\\
	T^3S & \frac{1}{8}\begin{pmatrix}4&2i\\ -2i&1\end{pmatrix}\\ 
	ST^2S & \frac{1}{4}\begin{pmatrix}0&0\\ 0&1\end{pmatrix}
	\end{array}
	\end{equation*}
  \caption{\label{table1}Generators of the orbit of the vacuum block.}
  \end{center}
\end{table}

Computing the sum of these six terms we find the correlation function
\begin{equation}\label{lister}
	G_{\text{candidate}}=\left|\calF^{\sigma\sigma\sigma\sigma}_\id\right|^2
+\frac{1}{4}\left|\calF^{\sigma\sigma\sigma\sigma}_\epsilon\right|^2,
\end{equation}
where the overall normalisation is fixed by demanding that the OPE with the identity is unity, for example from the behaviour as $x\to 0$. 

Equation \rref{lister} is precisely the correct answer for the $\sigma\sigma\sigma\sigma$ correlation function. 
We note that the modular average has correctly determined that $\sigma\sigma$ fuses only to the identity and $(h,\bar{h})=(\tfrac12,\tfrac12)$ scalars, and given the relative coefficient of the $\left|\calF^{\sigma\sigma\sigma\sigma}_\id\right|^2$ and $\left|\calF^{\sigma\sigma\sigma\sigma}_\epsilon\right|^2$ terms.  This coefficient is the three point function coefficient
\be
C_{\sigma\sigma\epsilon}^2=\frac{1}{4},
\ee
which is the only non-trivial three point coefficient for the Ising model.  In this way we see that the modular average has exactly reproduced the Ising model three point coefficients, taking only the central charge and dimensions of $\sigma$ as inputs.

\paragraph{Mixed four-point functions:}\mbox{}\\

We now consider the $\sigma\sigma\epsilon\epsilon$ four-point function, the only nontrivial example for which the vacuum block is exchanged in some channel. One might expect that we need to consider a nine-dimensional space of blocks, for the three operator dimensions that are exchanged and the three independent permutations of operators, giving the S,T and U channel for the exchange. However, in this case, only three of these blocks are produced as modular images of the identity exchange. These are one block for each channel, with the unique exchange allowed by the Ising fusion rules.

Explicitly, casting the results of \cite{francesco2012conformal,alvarez1989topics,Dotsenko:1985hi,Dotsenko:1984ad,Dotsenko:1984nm} in our conventions, the relevant blocks are
\begin{align}
\calF^{\sigma\sigma\epsilon\epsilon}_{\id}(x)&=\frac{1-\frac{x}{2}}{(1-x)^{5/16} x^{3/8}} \nonumber\\
\calF^{\sigma\epsilon\epsilon\sigma}_\sigma(x)&= \frac{1+x}{(1-x)^{3/8} x^{5/16}}\\
\calF^{\sigma\epsilon\sigma\epsilon}_\sigma(x)&= 
\frac{1-2 x}{((1-x) x)^{5/16}}.\nonumber
\end{align}
It is easy to see from these expressions that the $\Gamma(2)$ subgroup leaving the operators in the original order, generated by monodromies around $x=0$ ($T^2$) and around $x=1$ ($ST^2S$), act on the blocks with a phase only. In the basis
\begin{equation}
	\vec{\calF}=\left\{
\calF^{\sigma\sigma\epsilon\epsilon}_\id\,,
\calF^{\sigma\epsilon\epsilon\sigma}_\sigma
\,
\calF^{\sigma\epsilon\sigma\epsilon}_\sigma
 \right\}\, ,
\end{equation}
the generators of the full modular group act as
\begin{equation}
	T=\begin{pmatrix}
 (-1)^{13/8} & 0 & 0 \\
 0 & 0 & (-1)^{27/16} \\
 0 & (-1)^{27/16} & 0
\end{pmatrix}
, \quad
S=
\begin{pmatrix}
 0 & 2 & 0 \\
\frac{1}{2}  & 0 & 0 \\
 0 & 0 & -1 \\
\end{pmatrix}
.
\end{equation}
The vacuum block in the S-channel, represented by the matrix with one in the top left and zero elsewhere, is left invariant under the subgroup $\Gamma(2)$, as well as under $T$, which together generate the index three congruence subgroup $\Gamma_1(2)$. The modular sum therefore has three terms, and each term reproduces the correlation function, expanded in different channels (with only one block appearing in each channel):
\begin{equation}
	\langle\sigma\sigma\epsilon\epsilon\rangle = \left|\calF^{\sigma\sigma\epsilon\epsilon}_\id\right|^2;\: \langle\sigma\epsilon\epsilon\sigma\rangle = \frac{1}{4}\left|\calF^{\sigma\epsilon\epsilon\sigma}_\id\right|^2;\: \langle\sigma\epsilon\sigma\epsilon\rangle = \frac{1}{4}\left|\calF^{\sigma\epsilon\sigma\epsilon}_\id\right|^2.
\end{equation}
This again reproduces the correct OPE coefficient $C_{\sigma\sigma\epsilon}^2=1/4$, as well as the fact that this is the only nontrivial fusion.

The above computation can be recast in the language of representations of the anharmonic group, as described in \cref{sec:crossingmodular}.  Since we have a pair of identical operators, the computation will involve only the trivial representation and one copy of the standard representation of $S_3$.

\subsubsection{Tricritical Ising model}
Next in the unitary series is the tricritical Ising model, with $c=7/10$.  The primary operators and dimensions are 
\begin{align}
	\id :\: h_{(1,1)}=h_{(3,4)}=0,\quad &\epsilon:\: h_{(1,2)}=h_{(3,3)}=\frac{1}{10},\quad \epsilon':\: h_{(1,3)}=h_{(3,2)}=\frac{3}{5}, \\
	\epsilon'':\: h_{(1,4)}=h_{(3,1)}=\frac{3}{2},\quad &\sigma':\: h_{(2,1)}=h_{(2,4)}=\frac{7}{16},\quad \sigma:\: h_{(2,2)}=h_{(2,3)}=\frac{3}{80}.\nonumber
\end{align}

The four-point functions of all five nontrivial scalars of the model are reproduced from a modular sum with the Virasoro vacuum block as a seed. This is trivial for $\epsilon''$, since the fusion rule $\epsilon''\times\epsilon''=\id$ implies that the vacuum block alone is modular invariant, and the $\sigma'$ case is very similar to the case of $\sigma$ in the Ising model. For the other three, the sum does not truncate and we need to include an infinite number of terms\footnote{The representation of $PSL(2,\ZZ)$ for the four-point function of $\epsilon'$ is the same (up to a phase) as that of $\epsilon$. Although $\epsilon'$ is a third-order operator, the fusion rule $\epsilon'\times\epsilon'=\id+\epsilon'$ implies that the modular group acts invariantly on the two-dimensional space spanned by the $\id$ and $\epsilon'$ blocks. Thus, the analysis of four-point function of $\epsilon$ is the same as that of $\epsilon'$, consistent with the fact that $C_{\epsilon\epsilon\epsilon'}^2=C_{\epsilon'\epsilon'\epsilon'}^2$. This can be understood as a consequence of the model being secretly supersymmetric, with $\epsilon$ and $\epsilon'$ in the same supermultiplet \cite{Friedan:1984rv,Qiu:1986if}.}.

Modular-averaging four-point functions of identical operators gives all OPE coefficients of the form $C_{\op_a \op_a \op_b}$. The others can be obtained by considering correlation functions of two pairs of identical operators $\langle \op_a \op_a\op_b \op_b\rangle$. In the tricritical Ising model, the modular average of the vacuum block reproduces all OPE coefficients by studying these two types of four-point functions. We note that for other mixed correlators, for example $\langle \op_a \op_a \op_a \op_b \rangle$, the identity operator does not appear in the decomposition in terms of the blocks in any channel. In these cases, the light ``seed'' may be taken as the lightest operator appearing in any of the three channels.

\paragraph{Four-point function of $\sigma'$}\mbox{}\\

For the $\sigma'$-four-point function, the modular group acts on the subspace spanned by the vacuum block and the $\epsilon''$ exchange block, with the generators acting as
\begin{equation}
	T = e^{-7 i \pi/12 } \begin{pmatrix}1&0\\0&-i
	\end{pmatrix},\quad
	S=\frac{1}{\sqrt{2}}\begin{pmatrix}
		1&\frac{8}{7}\\ \frac{7}{8}&-1
	\end{pmatrix}
\end{equation}
taking the result from the appendix. The modular images of the vacuum block generate six distinct terms, and after normalisation this gives the correlation function
\begin{equation}
	G_{\text{candidate}}=
\left|\calF_\id\right|^2
+\frac{49}{64}\left|\calF_{\epsilon''}\right|^2.
\end{equation}
This is the correct correlation function, and gives the right value of the OPE coefficient $C_{\sigma'\sigma'\epsilon''}$.

\paragraph{Four-point function of $\epsilon$}\mbox{}\\

For the $\epsilon$ four-point function, the action of the modular group on the vacuum block generates only one other internal dimension, from the $\epsilon'$ block. The representation acting on this is
\begin{equation}
	T=e^{-2i\pi/15}\begin{pmatrix}
		1 & 0 \\
 0 & e^{3 i \pi/5}
	\end{pmatrix},\quad
	S= \begin{pmatrix}
		\frac{\sqrt{5}-1}{2}& \frac{\sqrt{5}-1}{2}\frac{\Gamma \left(\frac{1}{5}\right) \Gamma \left(\frac{8}{5}\right)}{\Gamma \left(\frac{2}{5}\right) \Gamma \left(\frac{7}{5}\right)} \\
		\frac{\Gamma \left(\frac{2}{5}\right) \Gamma \left(\frac{7}{5}\right)}{\Gamma \left(\frac{1}{5}\right) \Gamma \left(\frac{8}{5}\right)} &-\frac{\sqrt{5}-1}{2}
	\end{pmatrix}.
\end{equation}
In this case, the orbit of the vacuum block under the modular group appears to be infinite, so the sum does not truncate. We can, however, compute the sum numerically and check that it converges to the correct OPE\footnote{The sum does not converge in the standard sense, since the individual terms do not tend to zero. However, since we are normalising the result by an overall factor in the end, we can proceed as follows: we first choose an order for the terms, normalise the partial sums by an appropriate factor, and take the limit as we add more and more terms in the chosen order.  We expect that an unfortuitous choice of order could lead to any answer, as for conditionally convergent sums, but the hope is that any natural choice of ordering gives the same finite answer.}.

\begin{figure}
\centering
  \includegraphics[width=0.7\linewidth]{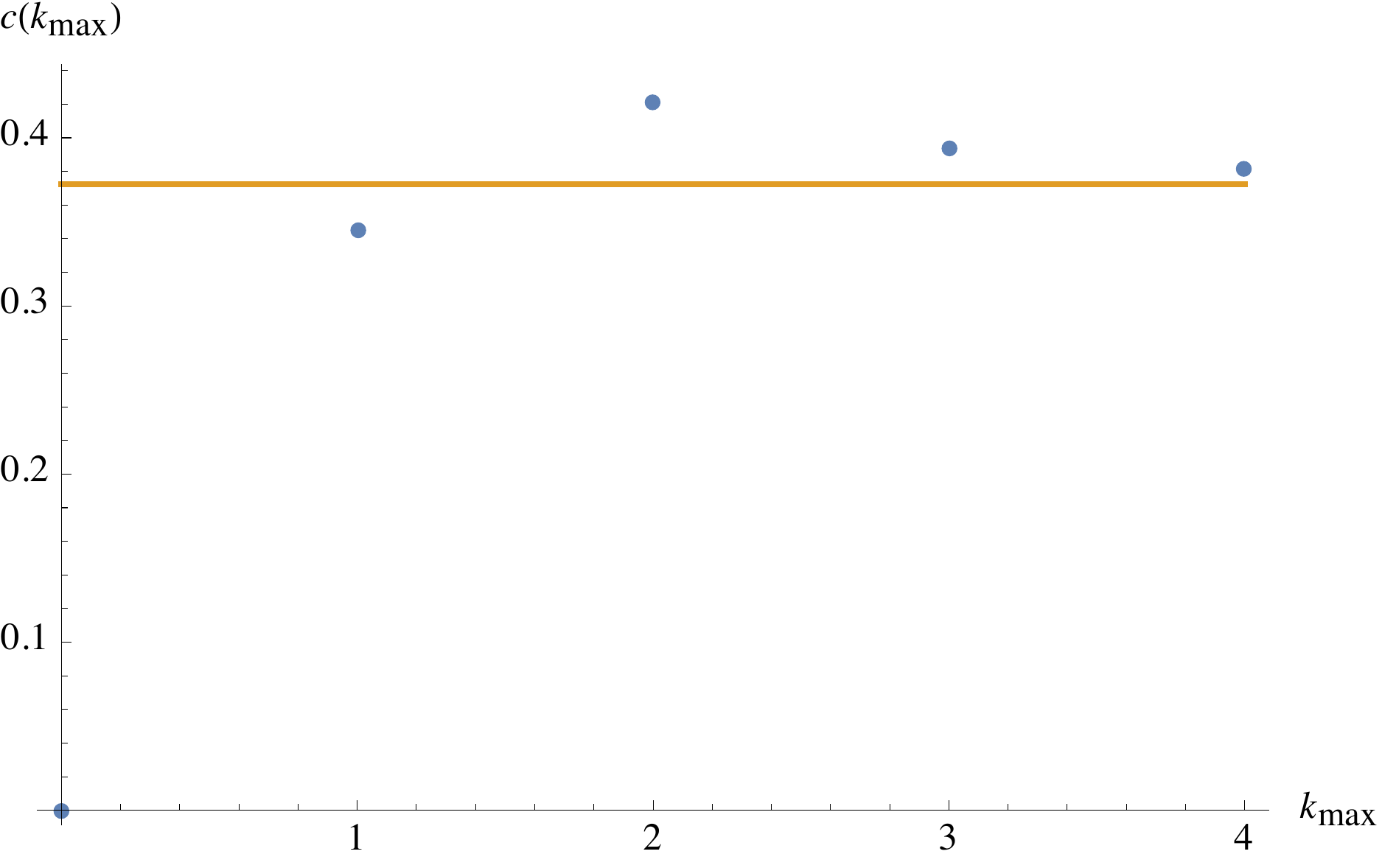}
  \caption{The blue dots show $c(k_{max})$, plotted agains $k_{max}$. 
  The orange line on the RHS is approximately $0.372$ -- the exact OPE coefficient. 
  The last point on the RHS is approximately $0.381$ which is about $2\%$  from $0.372$.}
  \label{fig:fig1}
\end{figure}

For our numerical checks, we performed the sum over the distinct images produced by products of $S$ and $T$ acting on $\calF_\id$, organised by the number of generators in the element of the modular group. Specifically, defining the length of a generator $\gamma$ as the minimal $k$ such that we may write
\begin{equation}
	\gamma = S^{m_1} T^{n_1}S^{m_2}T^{n_2}\cdots S^{m_k}T^{n_k}
\end{equation}
for  $m_i=0,1$ and non-negative integers $n_i$ up to the order of $T$, we sum over all distinct images of the seed, taking words of length at most $k_\mathrm{max}$:
\begin{align}
G_{candidate}&= N(k_{max})^{-1}\left. \sum_{\mathrm{length}(\gamma)\leq k_\mathrm{max}}
 \left|\calF_\id(\gamma \tau)\right|^2 \right|_{\text{distinct}}
\nonumber\\
&=
\left|\calF_\id\right|^2
+(b(k_\mathrm{max})\calF_\id \calF_{\epsilon'}^*+b(k_\mathrm{max})^* \calF_\id ^* \calF_{\epsilon'})
+c(k_\mathrm{max})\left|\calF_{\epsilon'}\right|^2
\, .
\end{align}
Here the coefficients of the blocks, after normalising the identity contribution to unity, are given by a complex number $b(k_{max})$ and a real number $c(k_{max})$. The subscript `distinct' in the sum is to indicate that we are only summing over distinct terms.
With this method, taking $k_{max}$ up to 4, which generates approximately $10^6$ distinct terms in the sum, the numerical results are consistent with the sum reproducing the correct OPE coefficients.
The off-diagonal terms are small, with $|b(k_{max}=4)|\approx 10^{-9}$, while for $c(k_{max})$ the result is around 2\% from the known exact value, and approaching it as more terms are added as shown in \cref{fig:fig1}:
\begin{equation}
	\left. G_\mathrm{candidate} \right|_{k_{max}=4}
\approx
\left|\calF_\id\right|^2
+0.381\left|\calF_{\epsilon'}\right|^2
+O(10^{-9}).
\end{equation} 

\subsubsection{The diagonal $M(12,11)$ minimal model}
\label{eq:M1211}
Unitary minimal models sometimes have Virasoro scalar primaries with even-integer dimension. 
In this case the $T$-matrix will have repeated eigenvalues, and hence invariance under $T$ does not require the four-point function to be a diagonal sum of conformal blocks squared, so non-scalar operators may be exchanged. In this case, there may be more than one solution to the crossing equations\footnote{See \cite{fuchs1989computation,fuchs1989operator} for discussions related to non-uniqueness of solutions to crossing equations.}. On the other hand, the modular sum of the vacuum block yields a unique crossing-symmetric answer.  We can then ask whether the modular average of the vacuum block reproduces the correct three point coefficients.

We will  answer this question in the $M(12,11)$ unitary minimal model, which has central charge $21/22$.  The diagonal model can be realised as the $\widehat{su}(2)$ coset model\footnote{For detailed discussion, see Chapter 18.3 of \cite{francesco2012conformal}.} 
\begin{equation}
	\frac{ \widehat{su}(2)_k \oplus \widehat{su}(2)_1}{\widehat{su}(2)_{k+1}}
\end{equation}
at level $k=9$, with $55$ Virasoro scalar primaries.

We will focus on the $\phi_{(1,4)}$ operator, which fuses with itself to four primaries
\begin{equation}
\phi_{(1,4)}\times\phi_{(1,4)}=\id+\phi_{(1,3)}+\phi_{(1,5)}+\phi_{(1,7)}
\end{equation} with dimensions
\begin{equation}
h_{(1,4)}=31/16,\quad
h_{(1,3)}=5/6,\quad
h_{(1,5)}=7/2,\quad
h_{(1,7)}=8,
\end{equation}
so, in particular, $h_{(1,7)}$ is an even integer.

\paragraph{Four-point function of $\phi_{(1,4)}$}\mbox{}\\
 The $T$ matrix corresponding to the four point function of $\phi_{(1,4)}$ is
\begin{equation}
T=e^{-\frac{4}{3} \pi i h_{(1,4)} }
\begin{pmatrix}
1 &0&0&0 \\
0& e^{\pi i h_{(1,3)}}&0&0 \\
0 & 0& e^{\pi i h_{(1,5)}}&0 \\
 0 & 0&0& e^{\pi i h_{(1,7)}}
\end{pmatrix}
=e^{-31 \pi i/12 }
\begin{pmatrix}
1 & 0&0&0 \\
0 & e^{5\pi i/6}&0&0 \\
0 & 0& -i&0 \\
 0 & 0&0& 1
\end{pmatrix}.
\end{equation}
  Invariance under $T$ then allows for off-diagonal terms in the corners of the $C$ matrix, restricting the form of the four point function to be
\begin{equation}
G=\left|\calF_\id\right|^2
+C_1\left|\calF_{(1,3)}\right|^2
+C_2\left|\calF_{(1,5)}\right|^2
+C_3\left|\calF_{(1,7)}\right|^2
+D_1 \calF_\id\calF_{(1,7)}^*
+D_2 \calF^*_\id \calF_{(1,7)}.
\end{equation}
The $S$ matrix is rather complicated analytically, but its numerical value is\footnote{For this fourth-order correlator, we used the Mathematica codes in \cite{Esterlis:2016psv} to obtain the $S$ matrix. Note that our matrix $S$ is the transpose of that in \cite{Esterlis:2016psv}, as explained previously in \cref{footnote1}.}
\begin{equation}
S\approx
\begin{pmatrix}
 0.2989 & 0.1863 & 0.0922 & 1.3807 \\
 1.3098 & 0.5176 & 0. & -4.43 \\
 3.6153 & 0. & -0.7071 & 6.1137 \\
 0.2414 & -0.1102 & 0.0273& -0.1094 \\
\end{pmatrix}
.
\end{equation}

Imposing $S$-invariance allows a one-parameter family of crossing invariant solutions:
\bea
G&\approx&\left|\calF_\id\right|^2
+\left(7.031\, -23.7794 D_1\right)\left|\calF_{(1,3)}\right|^2
+\left(39.2118+66.3091 D_1\right)\left|\calF_{(1,5)}\right|^2
\nonumber\\
&&
+\left(0.1749\, -0.2957 D_1\right)\left|\calF_{(1,7)}\right|^2
+D_1 \left(\calF_\id \calF_{(1,7)}^*+ \calF^*_\id\calF_{(1,7)}\right).
\eea  The diagonal model corresponds to $D_1=0$.

Similar to the previous section, the modular sum over the vacuum block can be done numerically.  We have performed the sum up to $k_{max}= 5$, which generates approximately $10^4$ distinct terms in the sum. The result is that the modular sum yields the modular invariant OPE with $D_1\approx 0.12$ and, in particular, none of the OPE coefficients implied by $G_\text{candidate}$ converge to zero.

Given this, if we wish for the modular sum to produce the correct correlation function, it must be in a model whose spectrum contains the $\phi_{(1,4)}$ scalar, as well as all the operators appearing in the conformal block expansion: the $\phi_{(1,3)},\phi_{(1,5)}$ and $\phi_{(1,7)}$ scalars, as well as the chiral $\phi_{(1,7)}$ spin 8 current.

There are three modular invariant spectra for the $M(12,11)$ minimal model, corresponding to $(A_{10},A_{11})$, $(A_{10},D_7)$ and $(A_{10},E_6)$ in the ADE classification of minimal model spectra \cite{cappelli1987modular,cappelli1987ade,kato1987classification} (reviewed in \cite{francesco2012conformal}), these being the pairs of simply laced root systems with dual Coxeter numbers $(11,12)$. The $A_{11}$ model is the diagonal one, containing only scalars, so in particular does not have the spin 8 current. The current is also absent in the $D_7$ model, which in addition lacks a $\phi_{(1,4)}$ scalar, so this correlation function is not even part of that theory. Finally, the $E_6$ model has a spin 8 current in the spectrum, but no $\phi_{(1,3)}$ scalar. There is therefore no model containing all the required operators for this correlation function to appear in a modular invariant theory.

This example illustrates that if one takes only the modular average of the vacuum block one will not always correctly reproduce all three point coefficients. This may be improved by adding more information to the seed before performing the sum: for example, the correct correlation function for the diagonal model can be obtained in this instance by including the correct OPE coefficient for the $h=8$ scalar in the seed, as well as the vacuum contribution.

\subsection{A group theoretic perspective}

We will now describe a somewhat more mathematical reformulation of the above discussion.  This will motivate a redefinition of the modular average, allowing it to be computed more rigorously and systematically for infinite sums.

Abstractly, we can formulate our problem as follows.  We have the vector space $V$ spanned by conformal blocks, on which a group $\Gamma$ acts in some representation $R$ ($\Gamma$ is $PSL(2,\ZZ)$ if it acts faithfully, modulo the kernel of the representation if not). In the two-dimensional case, $V=B\otimes \bar{B}$, and $R$ is the tensor product of a representation on $B$ and its conjugate. A crossing symmetric correlation function is a vector $v\in V$ which is invariant under the action of $\Gamma$. Our strategy is to start with a choice of `seed' vector $v_0\in V$ (the vacuum block in the above minimal examples) and to sum over all its images in $\Gamma$:
\begin{equation}
	v\propto\sum_{\gamma\in\Gamma} R(\gamma)v_0.
\end{equation}
Note that (assuming for now that all the relevant sums converge) the dependence on the seed $v_0$ factors out, so we can solve the problem by finding the linear map $P_R$ associated to the representation $R$ defined by
\begin{equation}
	P_R\propto \sum_{\gamma\in\Gamma} R(\gamma).
\end{equation}
We will show that for finite groups, $P_R$ is a projection canonically associated to the representation $R$.  We can characterise this projection more generally, including cases of relevance in our discussion where the convergence is less obvious. This could be regarded as an alternative proposal to construct correlation functions, motivated by the sum over $\Gamma$, and equivalent in many cases, but rigorously defined and sometimes more easily calculable.

Let us begin by taking $\Gamma$ to be a finite group\footnote{More generally, we could take $\Gamma$ to be compact, with Haar measure $\mu$, and define the average over the group as $\frac{1}{\mu(\Gamma)}\int_\Gamma d\mu$. Then the following discussion is essentially unaltered.} so the sum unambiguously makes sense. Now we can make use of the following standard results in the representation theory of finite groups:
\begin{itemize}
	\item Every finite-dimensional representation is equivalent to a unitary representation.
	\item Every finite-dimensional unitary representation is completely reducible (i.e.\ it decomposes as a direct sum of irreducible representations).
	\item The \emph{grand orthogonality theorem}, which states that for irreducible unitary representations $R_1,R_2$, the sums over $\Gamma$ of matrix elements are orthonormal, in the sense that
		\begin{equation}
			\frac{1}{|\Gamma|}\sum_{\gamma\in\Gamma} R_1(\gamma)_{i j}^*R_2(\gamma)_{i' j'} = \begin{cases}
				0 & R_1,R_2 \text{ inequivalent} \\
				\frac{1}{\dim(R_1)}\delta_{i i'}\delta_{j j'} & R_1=R_2 
			\end{cases}
		\end{equation}
\end{itemize}
As an immediate corollary of this last statement, choosing $R_1$ to be the trivial representation, we find that the sum over the group of a matrix element of a nontrivial irreducible unitary representation vanishes.

From the first two of the quoted results, we may choose a basis in which $R$ is block-diagonal, with each block being an irreducible unitary representation. There are a number of trivial representations appearing in this decomposition, and the subspace spanned by these representations is exactly the subspace of $V$ left invariant under the group action; it follows that the image of $P_R$ must be contained in this subspace. The orthogonality theorem then implies that in this basis where the representation is unitary, $P_R=\frac{1}{|\Gamma|}\sum_\Gamma R(\gamma)$ is the diagonal matrix with ones on the diagonal where the trivial representations live and zeroes elsewhere. More abstractly, the conclusion can be simply stated: \emph{There is an inner product on $V$ which is invariant under the action of $\Gamma$. The sum over $\Gamma$ is equivalent to the orthogonal projection, with respect to this inner product, onto the invariant subspace of $V$.} This inner product is not quite unique (there is a $GL(k)/U(k)$ choice for each inequivalent irreducible representation appearing $k$ times, for example, an overall scale if $k=1$), but the projection is independent of which is chosen. In particular, this definition of $P_R$ makes sense for any group, as long as $R$ is equivalent to a unitary representation.

We can refine this discussion further in the case of 2D CFTs using the additional structure implied by the factorisation of the blocks, so $V=B\otimes\bar{B}$. The action is then by conjugation, so $R$ is the tensor product of some representation $R_0$ on $B$ and its conjugate $R_0^*$ (or equivalently the dual of $R_0$ if it is a unitary representation). In this product, the identity representations in $R$ appear in a simple way when $R_0$ is unitary. This is because Schur's lemma implies that for unitary irreps $R_1$ and $R_2$, the identity appears in the decomposition of the tensor product $R_1 \otimes R_2^*$ into irreps exactly once when $R_1$ and $R_2$ are equivalent, and not at all when they are inequivalent. So writing $R_0 = \oplus_i k_i R_i$, where $R_i$ denote inequivalent irreps and $k_i$ their multiplicities, the dimension of the invariant subspace is $\sum_i k_i^2$.

When we have this decomposition into irreducible representations, after changing to the basis where $R_0$ is block diagonal, the projection acts on the matrix $C$ in a simple way. The elements of a block corresponding to inequivalent irreps acting from the left and right gets set to zero, while the blocks with the same representation acting on both sides (appearing on the diagonal in particular, and off the diagonal when there are multiple copies of the same representation) get projected to a multiple of the identity in that block, while preserving the trace.

The crucial requirement is that the representation $R_0$ is equivalent to a unitary representation. This is always true when $\Gamma$ is a finite group. Even in the infinite case there may exist a basis in which the representation is unitary.
Indeed, we will show that this is always the case when one considers 
identical operators in unitary minimal models, and restricts to the minimal subspace of exchange operators generated by the action on the vacuum block. We will give examples to show that relaxing either assumption may lead to a representation which is not equivalent to a unitary one.

To see this, note that when the group acts by conjugation, the solutions to crossing satisfy
\begin{equation}
	\gamma\cdot C\cdot \gamma^\dag =C\quad \forall\gamma\in\Gamma
\end{equation}
which means that $C$ is a Hermitian form on $B$, invariant under the action of $\Gamma$, with Hermiticity of $C$ guaranteed by reality of the correlation function. The only basis-independent information in such a form is its signature, the number of positive, negative and zero eigenvalues.  Thus there is a basis where the form has only $1,-1$ and $0$ along the diagonal. If there exists a form with definite signature, i.e.~with all eigenvalues having the same sign, then in the basis where the form is proportional to the identity, the invariance under $\Gamma$ is equivalent to unitarity of the representation. Then we may follow the logic of the above, decomposing $R_0$ into unitary irreps, and projecting onto the invariant subspace.

This happens for correlation functions of identical scalars in unitary minimal models, where we include only a subset of the exchange operators. This is because there is always a positive definite crossing-invariant solution, given by the diagonal minimal model, with squares of OPE coefficients of the scalars coupling to the external operator along the diagonal. Unitarity guarantees that the OPE coefficients are real, so there are no negative eigenvalues, and including only the internal operators appearing when the external operator fuses with itself (a subspace guaranteed to be invariant under $\Gamma$) ensures that there are no zero eigenvalues. Note that this only guarantees that the representation is unitary, not that the projection of the vacuum block will reproduce the diagonal model OPE coefficients, as can be seen from the $M(12,11)$ example.

\subsubsection{Example 1: Ising $\langle\sigma\sigma\sigma\sigma\rangle$}

By inspection of the representation relevant to the four-point function of $\sigma$ in the Ising model in \cref{isingsigmarep}, by rescaling to the basis $\{ \calF_\id,\calF_{\epsilon}/2\}$ the representation becomes unitary:
\begin{equation}
	T=e^{-i\pi/12}\begin{pmatrix}1&0\\0&i\end{pmatrix},\quad
	S=\frac{1}{\sqrt{2}}\begin{pmatrix}1&1\\1&-1\end{pmatrix}.
\end{equation}
This change of basis essentially amounts to absorbing the OPE coefficient into the block. This representation is irreducible, so there is a unique (up to multiples) invariant correlation function.

To show how the action on the two-dimensional space $B$ extends onto the four-dimensional space $V=B\otimes \bar{B}$ by conjugation, write the matrix on which it acts in the basis consisting of the identity and three Pauli matrices. Then the representation acts as
\begin{equation}
	T=\begin{pmatrix}
		1&0&0&0\\0&0&-1&0\\0&1&0&0\\0&0&0&1
	\end{pmatrix},\quad
	S=\begin{pmatrix}
		1&0&0&0\\0&0&0&1\\0&0&-1&0\\0&1&0&0
	\end{pmatrix}
\end{equation}
with the trivial representation appearing in the upper left component (once, as expected), and an irreducible three-dimensional representation in a second block.

\subsubsection{Example 2: Four-point function of $\phi_{(1,4)}$ in $M(12,11)$}
Rescaling the basis vectors by absorbing OPE coefficients again makes the representation of the modular group unitary in this case, and in fact makes the $S$ matrix look much simpler:
\begin{equation*}
	\!\!\!\! T=e^{-31 \pi i/12 }
\begin{pmatrix}
1 & 0&0&0 \\
0 & e^{5\pi i/6}&0&0 \\
0 & 0& -i&0 \\
 0 & 0&0& 1
\end{pmatrix},\;
S=\begin{pmatrix}
	\frac{\sqrt{3}-1}{\sqrt{6}} & \sqrt{\frac{1}{\sqrt{3}}-\frac{1}{3}} & \frac{1}{\sqrt{3}} & \frac{1}{\sqrt{3}} \\
 \sqrt{\frac{1}{\sqrt{3}}-\frac{1}{3}} & \sqrt{2-\sqrt{3}} & 0 & -\sqrt{\frac{2}{3} \left(\sqrt{3}-1\right)} \\
 \frac{1}{\sqrt{3}} & 0 & -\frac{1}{\sqrt{2}} & \frac{1}{\sqrt{6}} \\
 \frac{1}{\sqrt{3}} & -\sqrt{\frac{2}{3} \left(\sqrt{3}-1\right)} & \frac{1}{\sqrt{6}} & \frac{\sqrt{3}-2}{\sqrt{6}}
\end{pmatrix}.
\end{equation*}
Now we still have the freedom of a unitary change of basis, while keeping the representation unitary, which we will use to show that this representation is reducible. With change of basis matrix
\begin{equation}
P=\begin{pmatrix}
	-\frac{1}{\sqrt{3}} & 0 & 0 & \sqrt{\frac{2}{3}} \\
 0 & 1 & 0 & 0 \\
 0 & 0 & 1 & 0\\
 \sqrt{\frac{2}{3}} & 0 & 0 & \frac{1}{\sqrt{3}}
\end{pmatrix},
\end{equation}
writing $\gamma'=P\gamma P^{-1}$, the generators become
\begin{equation}
\!\!\!\! T'=e^{-7\pi i/12}
	\begin{pmatrix}
		1&0&0&0\\0&e^{5\pi i /6}&0&0\\0&0&-i&0\\0&0&0&1
	\end{pmatrix},\:\:
S'=\begin{pmatrix}
 -\sqrt{2-\sqrt{3}} & -\sqrt{\sqrt{3}-1} & 0 & 0 \\
 -\sqrt{\sqrt{3}-1} & \sqrt{2-\sqrt{3}} & 0 & 0 \\
 0 & 0 & -\frac{1}{\sqrt{2}} & \frac{1}{\sqrt{2}} \\
 0 & 0 & \frac{1}{\sqrt{2}} & \frac{1}{\sqrt{2}}
\end{pmatrix}
\end{equation}
and after this change of basis, the identity block is represented by
\begin{equation}
C_\text{seed}' = P
\begin{pmatrix}
	1&0&0&0\\
	0&0&0&0\\
	0&0&0&0\\
	0&0&0&0
\end{pmatrix}
P^\dag=
	\begin{pmatrix}
		\frac{1}{3} & 0 & 0 & -\frac{\sqrt{2}}{3} \\
 0 & 0 & 0 & 0 \\
 0 & 0 & 0 & 0 \\
 -\frac{\sqrt{2}}{3} & 0 & 0 & \frac{2}{3}
	\end{pmatrix}.
\end{equation}
After the projection, this will turn into a matrix proportional to the identity in the upper left $2\times2$ block, and twice the identity in the lower right $2\times2$ block. After projecting and changing basis back to the original one, we find an exact result for the OPE coefficients coming out of the modular sum (without the work of performing any sum). Translating to the notation used in \cref{eq:M1211}, 
where the crossing-invariant correlation functions are parametrised by $D_1$, the OPE coefficient with the spin $8$ current, we obtain $D_1=7499023/63406080\approx0.118$, consistent with the truncated numerical sum. This provides good evidence that the sum, the way we have defined it, does indeed give the same result as the group-theoretic method.

\subsubsection{Example 3: Yang-Lee model}
To illustrate what happens when we relax the unitarity condition, consider the $M(5,2)$ minimal model, corresponding to the Yang-Lee edge singularity. This model has central charge $c=-22/5$, and one primary operator apart from the identity, the scalar $\Phi$ with $h=-1/5$. For the four-point function of $\Phi$, we can compute the action of the modular group on the blocks as before, finding
\begin{equation}
	T=e^{-2\pi i/5}\begin{pmatrix}
		1&0\\0& e^{-i\pi/5}
	\end{pmatrix},\quad
	 S=
 \begin{pmatrix}
		-\varphi &-\varphi/\alpha \\ 
		\alpha& \varphi
	\end{pmatrix},
\end{equation}
where $\varphi=\frac{1+\sqrt{5}}{2}$ is the golden ratio, and $\alpha =\frac{\Gamma\left(\frac{1}{5}\right)\Gamma\left(\frac{6}{5}\right)}{\Gamma\left(\frac{3}{5}\right)\Gamma\left(\frac{4}{5}\right)}$. This representation has a unique invariant hermitian form (up to scale), with the OPE coefficients $C^2_{\Phi\Phi\id}=1$ and  $C^2_{\Phi\Phi\Phi}=-\alpha^2/\varphi$
 on the diagonal. Since the nontrivial OPE coefficient is imaginary in this model, this form has indefinite signature.

\subsubsection{Example 4: Ising $\langle\epsilon\epsilon\epsilon\epsilon\rangle$}\label{Isingeeee}
We have already commented that the vacuum block alone is modular invariant for the four-point function of the $\epsilon$ operator in the Ising model, so no sum is required to find a solution to crossing, since $\epsilon\times\epsilon$ fuses only to the identity. Despite this, we may still consider the action of the modular group on both the identity and $\epsilon$ blocks to illustrate the general pattern. The representation is given by
\begin{equation}
T=	\begin{pmatrix}
	e^{-2\pi i/3}&0\\ 0& -1
\end{pmatrix},\quad
S= \begin{pmatrix}
	1&\frac{10\Gamma\left(\frac{2}{3}\right)^2}{9\Gamma\left(\frac{1}{3}\right)}\\
	0& -1
\end{pmatrix}
\end{equation}
which is reducible, but not completely reducible. This means, in particular, that it cannot be equivalent to a unitary representation, and indeed the only invariant Hermitian form is degenerate, with the $\epsilon$ exchanged block being a zero eigenvector.

\section{Semiclassical limit}
\label{sec:semiclassical}

We have motivated the main construction of the paper -- a candidate correlation function obtained by summing a conformal block over all channels --  as an abstract method for solving the constraints of crossing. We will now explain how the same construction follows naturally from considerations of a semiclassical gravity dual.  We will focus again on two dimensional CFTs, and consider gravity in three dimensions with some `heavy' bulk particles quantised via the worldline formalism. Correlation functions are then found by integrating over all possible worldlines of these particles.  In the semiclassical limit this is dominated by solutions to the classical equations of motion, including the backreaction of the particles on the geometry. The action of a classical solution (along with perturbative corrections) will compute a conformal block in the dual CFT, and different channels correspond to different classical solutions. The sum over channels is therefore the same as the sum over saddle points in the bulk path integral.

It is not obvious that the crossing images under $PSL(2,\ZZ)$ for a given conformal block should be in one-to-one correspondence with classical bulk solutions. We will focus on one example -- the four-point function of a $\Delta=c/16$ scalar -- where it is possible to classify all the classical solutions and make this correspondence explicit. This will also give us a bulk knot theoretic interpretation of the channels of the conformal block, in terms of `rational tangles', and make close contact with the older idea of the partition function Farey tail.

\subsection{The conformal block Farey tail as first-quantised gravity}
\label{sec:firstquantise}
We will consider a CFT in two dimensions with a semiclassical bulk dual. This means, in particular, that the central charge $c=\frac{3\ell_{AdS}}{2G_N}\gg 1$ is large so that the Planck length is small in AdS units. The spectrum of primaries in the theory must also be constrained (see e.g. 
\cite{Heemskerk:2009pn, ElShowk:2011ag, Hartman:2014oaa, Belin:2014fna,Haehl:2014yla}).  In particular, there are some `light' primaries, whose dimension does not scale with $c$, which are described by perturbative bulk fields\footnote{A single bulk field does not give rise to a single primary, but rather to a tower of primaries coming from multiparticle states.  In the language of large $N$ gauge theories, these are multi trace operators. }.  There are also heavy states with dimension of order $c$, but with $\Delta<c/12$.  These are dual to massive particles in the bulk, which backreact on the geometry to form conical defects.  Finally we have states with $\Delta>c/12$, corresponding to bulk black hole microstates, with asymptotic density of states given by the Cardy (or Bekenstein-Hawking) formula. 
	Our strategy will be to quantise the light bulk fields (including the graviton) as well as the heavy bulk particles by computing perturbatively a path integral with an appropriate action.  The contribution of the black hole states 	will then follow from a non-perturbative sum over bulk saddle points.  
	
This is most familiar in the black hole Farey tail \cite{Dijkgraaf:2000fq} (see also \cite{Maloney:2007ud}), where the partition function of the theory is computed by summing over topologically distinct saddle points.   We will start by briefly reviewing this construction, before turning to the analogous computation of correlation functions.

We begin with the computation of the partition function of a two dimensional CFT as a sum over all states, weighted by the Boltzmann factor. This may be organised into a sum over only primary operators, with contributions from descendants packaged into the characters $\chi_p$ of the Virasoro (or perhaps some other extended) algebra:
\begin{equation}
	\mathcal{Z}(\tau) = \sum_{\text{all states}} q^{L_0-\frac{c}{24}} {\bar q}^{{\bar L}_0-\frac{c}{24}} = \sum_{\text{all primaries}} \chi_p(\tau)\bar{\chi}_p(\bar{\tau}).
\end{equation}
We wish to compute this using a Euclidean bulk path integral. The path integral is over all bulk solutions whose boundary is a torus, the spatial circle times the Euclidean time circle.  In  semiclassical gravity this sum is dominated by a set of saddle points, which are the classical solutions of Einstein's equations with torus boundary \cite{Maldacena:1998bw}. The leading order contribution of each saddle point is the classical bulk action, with bulk loops around these solutions giving corrections perturbative in $1/c$. One solution is thermal AdS (pure Euclidean AdS with periodic identification of Euclidean time). The action and loop corrections around this solution are computed by the characters of light bulk fields.  The Virasoro character comes from the graviton loops and other light primaries give loops for the corresponding bulk fields \cite{Giombi:2008vd}.  The other classical solutions of pure gravity are given by modular transformations of thermal AdS.  The sum over saddles is therefore a sum over the modular group, with the summand being the total of the characters of light primaries:
\begin{align*}
	\mathcal{Z}(\tau) &= \sum_{\substack{\text{saddle}\\ \text{points}}} e^{-c\, S_\text{classical} + S_\text{one-loop} + \dots}\\
	&= \sum_{\gamma\in PSL(2,\ZZ)/\ZZ} \sum_{\substack{\text{light}\\ \text{primaries}}} \chi_p (\gamma \tau) \bar{\chi}_p (\gamma \bar{\tau}).
\end{align*}
At leading order the partition function will be dominated by the geometry with least action. This means that the leading order partition function has first order phase transitions (the Hawking-Page transition \cite{Hawking:1982dh} in this case) as $\tau$ varies and different saddle points exchange dominance.  This phase transition will be smoothed out at finite $c$.

Comparing the CFT and gravity results, we see that the contribution from heavy states is accounted for in gravity by the contribution of a different bulk saddle.  In other words, the heavy states come from the light states, but propagating in a different channel (i.e. around a different cycle on the boundary torus).  The partition function is constructed as a modular sum over the characters of the light spectrum only. By construction, this is modular invariant, though it may not decompose into a sum over characters with positive density of states \cite{Maloney:2007ud,Keller:2014xba}.

Our proposal is that essentially the same strategy should be used to study correlation functions, with the characters now being replaced by conformal blocks. For definiteness, let us consider a description of gravity in which heavy particles (that is, with $\Delta$ of order $c$) are `first quantised', in the worldline formulation. The perturbative path integral is therefore over configurations of light fields, as well as over heavy particle worldlines (including interactions where worldlines may split and join). The single-particle states of the massive bulk particles, which we will take to be scalars for simplicity, are dual to CFT primaries  with energy of order $c$, but less than $c/12$ above the vacuum. The correlation functions of the corresponding heavy primary operators are again given by a bulk path integral,  but now imposing the boundary conditions that an appropriate particle worldline ends on the boundary at the insertion point of the heavy operator.

In the large $c$ semiclassical limit, the path integral is dominated by classical solutions, including the backreaction from heavy particles.  Each heavy particle worldline contributes a factor of $m L$ to the action, where $m$ is the mass and $L$ is the (regularized) proper length of the worldline.  The heavy particles also back-react on the geometry, creating a conical singularity with deficit angle $2\pi(1-\alpha)$, where $\alpha=1-6m/c=\sqrt{1-24h/c}$, along its worldline. In many cases, the action of these solutions corresponds to the contribution to the correlation function from an appropriate semiclassical conformal block \cite{Jackson:2014nla,Fitzpatrick:2015zha,Fitzpatrick:2015dlt,Hijano:2015qja,Hijano:2015rla,Chang:2016ftb,Chang:2015qfa}. Bulk graviton loops around the solution contribute to the perturbative corrections (in $1/c\sim G_N$) to the semiclassical blocks.  Loops from other light bulk fields contribute to blocks where the corresponding light primaries are exchanged.

Once again, the full correlation function should be given as a sum over the contributions from all classical solutions. Schematically
\begin{equation}
	\langle\mathcal{O}\cdots\mathcal{O}\rangle = \sum_{\substack{\text{classical}\\ \text{solutions}}} \sum_{\substack{\text{light}\\ \text{primaries}}} F_p
\end{equation}
where $F_p$ denotes the appropriate conformal block, with the light operator $p$ exchanged\footnote{We will discuss examples where some classical solutions correspond to the exchange of a heavy particle dual to a bulk conical defect, represented by an internal worldline of this particle in the bulk. The loop corrections due to light bulk fields do not then literally correspond to a sum over blocks of light primaries.}.

In the case of the four-point function, this is exactly of the form of our proposed correlation function \cref{candidate}, provided we can show that the sum over classical solutions includes the modular sum over channels described in \cref{sec:poincaresum}. We will now explain how the family of classical solutions corresponding to the modular sum arises topologically in the sum over classical worldlines.  We then give an explicit example where we can show that there is a unique solution for the topological classes associated with the sum over channels, and no others.

Just as in the case of the partition function, in the semiclassical limit the correlation function will be dominated by a particular classical solution.  As we vary the moduli (in this case the cross-ratio) this gives rise to first-order phase transitions in correlation functions.  An example of this is the well-known exchange in dominance of Ryu-Takayanagi surfaces for the entanglement entropy of two intervals, which can be thought of as a formal limit of correlation functions of twist operators in cyclic orbifolds of the theory \cite{Calabrese:2009qy}.  At finite $c$ this phase transition will be smoothed out by the subleading ``instanton" corrections to the correlation function.

\subsection{Rational tangles and modular invariance}

We begin by describing in more detail the bulk interpretation of the different $PSL(2,\ZZ)$ channels which appear in our conformal block Farey tail.

\begin{figure}
\centering
\begin{subfigure}{.3\textwidth}
	\includegraphics[width=.9\textwidth]{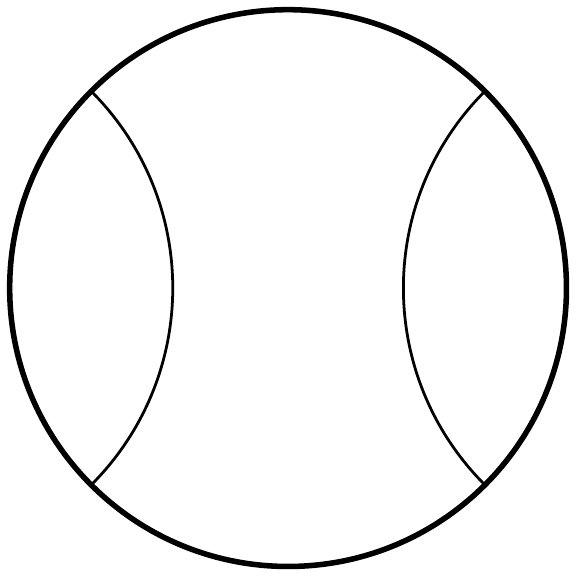}
	\caption{$t_\infty$\label{tinfty}}
\end{subfigure}
\begin{subfigure}{.3\textwidth}
	\includegraphics[width=.9\textwidth,angle=90]{infinityTangle}
	\caption{$t_0$\label{tzero}}
\end{subfigure}
\begin{subfigure}{.3\textwidth}
	\includegraphics[width=.9\textwidth]{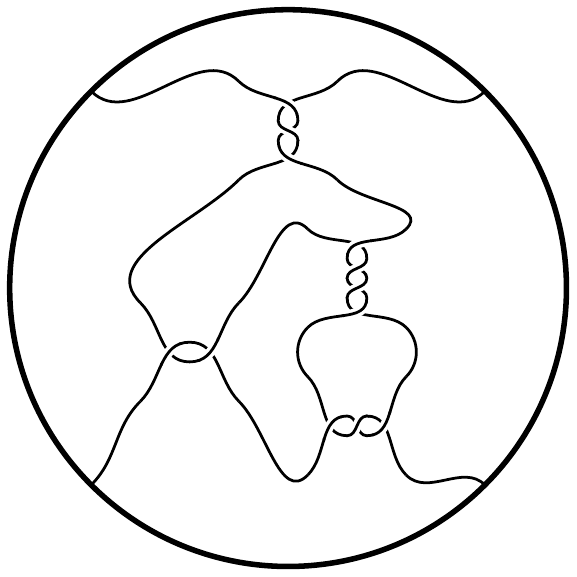}
	\caption{$t_{-29/74}$\label{tminus29over74}}
\end{subfigure}
	\caption{The rational tangles $t_\infty$, $t_0$ and $t_{-29/74}$.
	The last diagram should be compared with the continued fraction $-1/(3-1/(2-1/(-4-1/3)))=-\frac{29}{74}$.\label{ratTangles}}
\end{figure}

Consider the calculation of the Euclidean four-point function 
\begin{equation}
\langle\mathcal{O}(z_1)\mathcal{O}(z_2)\mathcal{O}(z_3)\mathcal{O}(z_4)\rangle
\end{equation}
of a heavy primary $\mathcal{O}$ which is dual to a massive bulk particle that sources a conical defect. One simple classical bulk
contribution to this correlator involves two bulk worldlines of this massive particle which join the $z_i$ on the boundary sphere in pairs.  In fact, there are many such contributions, with different topology.  A simple contribution, which we denote $t_\infty$ for reasons that will become clear below, is shown in \cref{tinfty}; it has one worldline joining $z_1$ and $z_2$ and another $z_3$ and $z_4$. This contribution is expected to dominate as we take the cross-ratio $x=\frac{(z_2-z_1)(z_4-z_3)}{(z_3-z_1)(z_4-z_2)}\to 0$.  This is the channel where we fuse $\mathcal{O}(z_1)$ with $\mathcal{O}(z_2)$ and $\mathcal{O}(z_3)$ with $\mathcal{O}(z_4)$.  The corresponding conformal block comes from the exchange of the identity operators and descendants, along with other light operators which would give additional loop corrections.

Suppose now that we begin with this solution as a function of the insertion points $\{z_i\}$. We may then generate further solutions by analytic continuation.  We continuously vary the $z_i$ without bringing two insertion points together, obtaining at the end the same configuration of points we started with, albeit with the $z_i$ possibly permuted. For example, we may start with $t_\infty$ and rotate the boundary sphere to cyclically permute the insertion points, obtaining the tangle $t_0$ shown in \cref{tzero}.  This configuration joins $z_1$ to $z_4$ and $z_2$ to $z_3$, corresponding to the T-channel, where we fuse $\mathcal{O}(z_1)$ with $\mathcal{O}(z_4)$ and $\mathcal{O}(z_2)$ with $\mathcal{O}(z_3)$, dominant as $x\to 1$.

\begin{figure}
\centering
	\begin{subfigure}{.3\textwidth}
	\includegraphics[width=.9\textwidth]{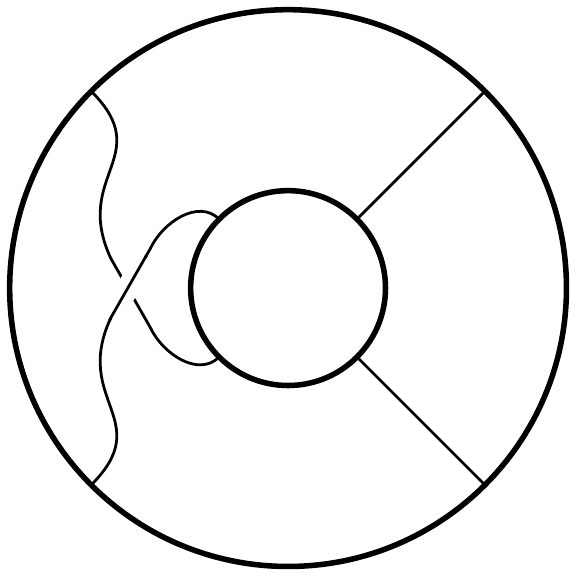}
	\caption{$\sigma_1$}
	\end{subfigure}
	\begin{subfigure}{.3\textwidth}
	\includegraphics[width=.9\textwidth,angle=90]{braidGen}
	\caption{$\sigma_2$}
	\end{subfigure}
	\begin{subfigure}{.3\textwidth}
	\includegraphics[width=.9\textwidth,angle=180]{braidGen}
	\caption{$\sigma_3$}
	\end{subfigure}\\
	\vspace{7pt}
	\begin{subfigure}{.6\textwidth}
		$\vcenter{\hbox{\includegraphics[width=.45\textwidth]{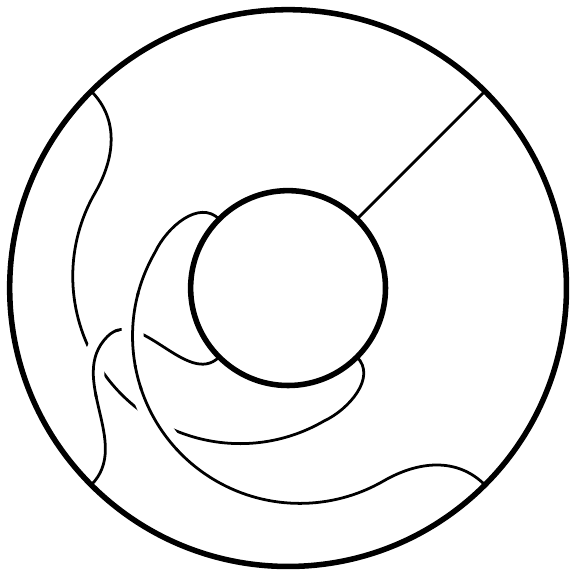}}}\,\scalebox{1.5}{=}\,\vcenter{\hbox{\includegraphics[width=.45\textwidth]{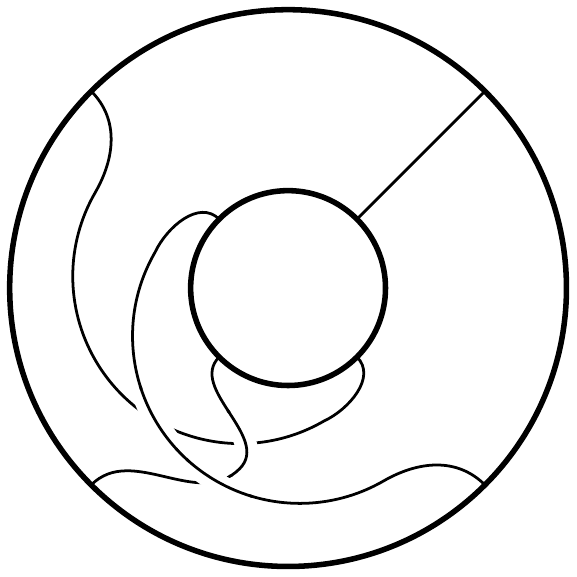}}}$
		\caption{$\sigma_1\sigma_2\sigma_1=\sigma_2\sigma_1\sigma_2$}
	\end{subfigure}
	\begin{subfigure}{.3\textwidth}
		\includegraphics[width=.9\textwidth]{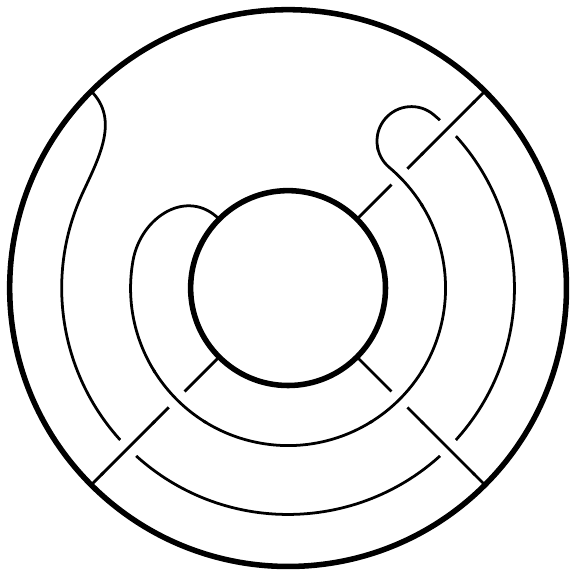}
		\caption{$\sigma_1\sigma_2\sigma_3^2\sigma_2\sigma_1=1$}
	\end{subfigure}
	\caption{The spherical braid group generators and relations:
		$B_4(S^2)$ is generated by $\sigma_1,\sigma_2,\sigma_3$, with the relations $\sigma_1\sigma_3=\sigma_3\sigma_1$, $\sigma_1\sigma_2\sigma_1=\sigma_2\sigma_1\sigma_2$, $\sigma_2\sigma_3\sigma_2=\sigma_3\sigma_2\sigma_3$, and $\sigma_1\sigma_2\sigma_3^2\sigma_2\sigma_1=1$. The outer in inner circles represent cross-sections through a two-sphere so, for example, the last braid is trivial because the strand can be unwrapped around the front and back of the internal $S^2$. \label{braid}
}
\end{figure}

In mathematical terms, the generation of further solutions by analytic continuation can be described as \emph{braiding} of the insertion points: the four-strand braid group on the sphere $B_4(S^2)$, described in \cref{braid}, acts on the space of solutions.

As we analytically continue the solutions, we expect the worldlines of the particles to stay apart: they do not intersect or pass through one another, and they do not split and join. This means that the worldlines can be usefully categorised by their topological class, defining what is known in knot theory as a `2-tangle'. An $n$-tangle is, roughly, a configuration of $n$ strings in the ball $B^3$ which end on $2n$ fixed boundary points, with configurations considered equivalent if and only if they can be continuously deformed into one another without strings passing through one another while leaving the boundary anchor points fixed. The braid group $B_{2n}(S^2)$ acts on the space of $n$-tangles in the obvious way.

The set of solutions we describe, obtainable from analytic continuation of $t_\infty$, gives only a limited set of topological classes of tangles.  We get the orbit of the 2-tangle $t_\infty$ under the braid group, which is known as the set of \emph{rational tangles}, denoted $\rats$.  Informally, $\rats$ is the set of tangles that can be untangled by moving the boundary anchoring points around on the sphere.  This excludes, for example, tangles with a strand that is by itself knotted in the bulk.   

\begin{figure}
\centering
	\includegraphics[width=.4\textwidth]{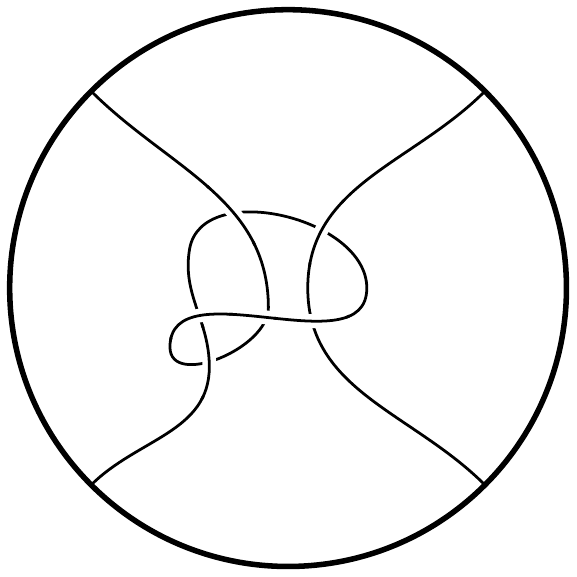}\qquad
	\includegraphics[width=.4\textwidth]{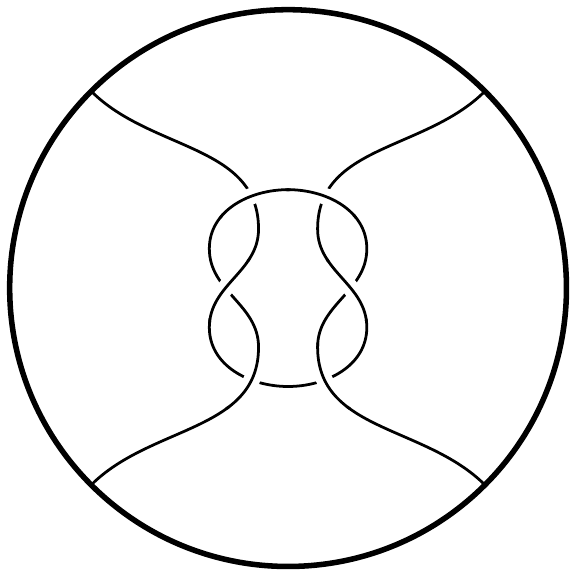}
	\caption{Examples of 2-tangles which are not rational, so may not be untangled only by moving the boundary points. No such worldline topologies appear as saddle points to the gravitational path integral.\label{nonRatTangles}}
\end{figure}

Rational tangles were classified by Conway\cite{conway1970enumeration}: they are in one-to-one correspondence with the rational numbers and infinity, $\rats = \{t_r|r\in \QQ\cup\{\infty\}\}$. A non-trivial rational tangle is shown in \cref{tminus29over74}. Examples of 2-tangles not included in this set are shown in \cref{nonRatTangles}.

To explain this classification, we need one small lemma about rational tangles. Drawing them as in \cref{ratTangles}, with the tangles anchored at diagonal points (traditionally labelled NE,SE,SW,NW with reference to points of the compass) in the plane of the page, the rational tangles are invariant under rotation by $\pi$ about any of the three axes running vertically, horizontally, and coming out of the page. This is easy to prove by induction: it is clearly true for $t_\infty$, and if it holds for some tangle, it is straightforward to verify that it continues to hold after acting with any of the three braid group generators. Notice that these rotations can be performed while keeping the cross-ratio of the boundary points constant.

As a consequence of the lemma, we learn that the braid group does not act faithfully: $\sigma_1\sigma_3^{-1}$ acts trivially on any rational tangle, since that element has the same effect as a $\pi$ rotation around the horizontal axis. To find the group that acts faithfully, we should therefore quotient the braid group by the normal subgroup generated by $\sigma_1\sigma_3^{-1}$. Write $T$ for the coset of $\sigma_1$ (or $\sigma_3$), and $S$ for the coset of $\sigma_1\sigma_2\sigma_3$, so that, in particular, $\sigma_2$ is in the coset $T^{-1}ST^{-1}$. Then the relations for the quotient can be written as
\begin{equation}
	B_4(S^2)/\mathcal{N}\langle\sigma_1\sigma_3^{-1}\rangle = \langle S,T|S^2=1,(ST)^3=1\rangle.
\end{equation}
These are the defining relations of the modular group $PSL(2,\ZZ)$. The action of $S$ and $T$ on tangles is simple, $S$ acting as a $\pi/2$ rotation of the knot diagram in the axis coming out of the page (in either sense, since a $\pi$ rotation acts trivially, so that $S=S^{-1}$), and $T$ by twisting the strands on the left, the same as $\sigma_1$ in \cref{braid}.

The set of rational tangles $\rats$ is given by the orbit of $t_\infty$ under the modular group. By the orbit-stabiliser theorem, this is the same as the set of cosets of the stabiliser, the subgroup leaving $t_\infty$ invariant. It is clear that $T$ acts trivially on $t_\infty$, and in fact the stabiliser is exactly the $\ZZ$ generated by that element (to show that the stabiliser is no larger is the nontrivial part of Conway's classification; a proof can by found in \cite{goldman1997rational}). We therefore have
\begin{equation}
	\rats \simeq \frac{PSL(2,\ZZ)}{\ZZ} \simeq \QQ\cup\{\infty\}.
\end{equation}
The last equality follows by considering the action of $PSL(2,\ZZ)$ on $\QQ\cup\{\infty\}$ by fractional linear transformations $r\mapsto\frac{a r+b}{c r+d}$; the stabiliser of $r=\infty$ is precisely the powers of $T$, which acts here by $T(r)=r+1$ (with $S(r)=-1/r$). This completes the classification. The action of the modular group on tangles allows one to simply describe the rational tangle $t_r$ in terms of the continued fraction decomposition of the rational number $r$, as noted in \cref{tminus29over74}.

We have learned that there is a natural action of the modular group on the space of rational tangles. This is the same as the action on conformal blocks. To see this, recall that the braids leaving all rational tangles invariant are precisely those that can be done while keeping the cross-ratio $x$ constant. Other braids will cause $x$ to traverse a topologically nontrivial path through cross-ratio space, ending at $x$ or one of its anharmonic images $1/x,1-x,\frac{1}{1-x},\frac{x}{x-1},1-\frac{1}{x}$ depending on how the operators are permuted. This precisely mirrors the discussion of \cref{sec:crossingmodular}, in which the different nontrivial paths correspond to distinct channels of a conformal block. We may pass to the universal cover of cross-ratio space, which is the upper half-plane, and the braids will there correspond to a path joining the initial $\tau$ to one of its images $\frac{a\tau+b}{c\tau+d}$ under the modular group. The braids corresponding to the generators act on $\tau$ in the usual way, $S\cdot\tau=-1/\tau$ and $T\cdot \tau=\tau+1$ . Finally, the tangle $t_\infty$ corresponds to the usual S-channel block, which is invariant under $T$ as required.

If the operators are not identical, it may be useful to distinguish between operator insertion points. From the cross-ratio point of view, this means we allow $x$ to continue only to a subset of its anharmonic images, so we reduce to a subgroup of the full modular group. From the $\QQ$ classification, the three ways to join the boundary points with tangles in pairs are distinguished by whether the numerator and denominator of the rational number are even or odd. In particular, if we require the tangles to be joined as in the original configuration of $t_\infty$, for example to compute the vacuum block of $\langle \mathcal{O}_1\mathcal{O}_1\mathcal{O}_2\mathcal{O}_2\rangle$, we must restrict to rational numbers with even denominator (such as $0$!) and odd numerator. For example, the tangle $t_{-29/74}$ in \cref{tminus29over74} has associated rational number of this form, and pairs the boundary points in the same way as $t_\infty$. This set is invariant under the congruence subgroup $\Gamma_1(2)$ ($a,d$ odd and $c$ even). If we distinguish all boundary points, we should consider not the braid group $B_4(S^2)$ acting on the boundary, but the pure braid group $P_4(S^2)$, the normal subgroup restricting to braids that do not permute the marked points. A similar analysis leads us to consider the congruence subgroup $\Gamma(2)$ ($a,d$ odd and $b,c$ even), which can be understood as the subgroup of the modular group that leaves all three possible ways of joining boundary points invariant when acting on tangles. This should be compared with the discussion of \cref{sec:crossingmodular}. The refined notion of vector valued modular functions used there can be realised for tangles by including a label on the endpoints of the strands, which allows us to act with the entire modular group while keeping track of the permutation of operators.

These topological considerations relate the crossing images of a conformal block to a particular set of bulk solutions. It is not clear that these are the only classical solutions, so the proposed sum over modular images of light blocks may miss some saddle points to the path integral. To help to justify this, we now give one example where we can prove that the rational tangle construction exhausts all solutions.

\subsection{The semiclassical $h=c/32$ conformal block}

We will now consider an example where all classical saddle points in the worldline formulation can be classified.  We consider the four-point correlation function of a dimension $h=\bar{h}=c/32$ scalar in a 2D CFT in the semiclassical limit. This example will also make a more direct connection to the partition function Farey tail.

We begin by considering the saddle points contributing to a four point function of a scalar of weight $h={\bar h}$ in the semiclassical limit.  In this limit, we need to compute the action of a pair of particles 
 which propagate through the bulk between the boundary operator insertion points. Each worldline contributions a factor $m L$, where $m$ is the mass of the bulk particle and $L$ is the (regulated) proper length of the worldline.  When the mass of the particle is of order the central charge, we must also include gravitational backreaction.  Each particle creates a delta-function source of stress-energy supported along the worldline, and Einstein's equations then imply that the worldline is replaced by a conical singularity.  The conical deficit angle is related to the mass of the particle by \be 2\pi(1-\alpha)=8\pi G_N m.\ee  In terms of the dimension $h$ of the operator we have 
 \begin{equation}
 	\alpha = 1-\frac{6m}{c} = \sqrt{1-\frac{24h}{c}}.
 \end{equation}
 Note that, since $0<\alpha<1$, the operator must have $0<h<\frac{c}{24}$.
 
 We must now find the gravitational action of the backreacted configuration of two worldlines, where no other particles are exchanged in the bulk.  This will give the leading semiclassical contribution to the vacuum conformal block in the channel where the pairs of boundary points joined by the particle worldlines fuse to the identity operator. Generically, the interaction between the two particles means that the geometry cannot be explicitly found.  Thus it is not possible to find a closed form expression for this semiclassical block.

However, at the special value $h=c/32$ the deficit angle is exactly $\pi$, which allows us to make progress.  The trick is to consider not the original geometry, but the twofold cover, branched along the particle worldlines. This solves the equations of motion, but it is smooth everywhere, since we do not have any other massive particle exchanged. The boundary geometry in this example is particularly simple, being a torus.  This is exactly the situation one encounters in the computation of four point functions of twist operators in a $\ZZ_2$ orbifold theory \cite{Lunin:2000yv}; indeed such operators have precisely dimension $h=c/32$. This arises also in the computation of the second R\'enyi entropy for a pair of intervals \cite{Headrick:2010zt,Calabrese:2009ez,Calabrese:2010he}.

We must begin by finding the smooth solutions to 3D gravity with torus boundary, which are known \cite{Maloney:2007ud} to be thermal AdS and the Euclidean BTZ black hole, and their `$SL(2,\ZZ)$ black hole' generalisations \cite{Maldacena:1998bw}, described in more detail below. In all of these solutions, the $\ZZ_2$ covering group of the boundary extends as an isometry into the bulk.  Taking the quotient by this $\ZZ_2$ gives the desired solutions with conical deficit worldlines. It follows that we have the complete classification of all such classical solutions.

As we will describe below, the solutions with torus boundary are labelled by an upper half-plane parameter $\tau$ (parameterising the conformal structure of the torus) as usual.   Images of $\tau$ correspond to different solutions with the same (or anharmonically related) cross-ratio.  Moreover, as described in the previous section, the topology of the conical defects is that of a rational tangle. 

\subsubsection{Gravity solutions\label{gravitySols}}

Let us now be more explicit about the classical solutions. The double cover of the boundary can be written as an elliptic curve
\begin{equation}
	y^2 = z(z-x)(z-1)
\end{equation}
where $z$ is the usual coordinate on the sphere, and $y$ picks up a sign after circling the branch points at $0,x,1,\infty$; this sign labels the two sheets of the cover. With the familiar description of the torus as the complex plane modulo a lattice ($u\in\CC$, with identifications $u\sim u+1\sim u+\tau$, for some $\Im\tau>0$), the map to the Riemann sphere giving $z$ in terms of $u$ is a doubly periodic function, which is essentially the Weierstrass $\wp$-function (up to some M\"obius map). This map is two-to-one, mapping $u$ and $-u$ to the same $z$, excepting at the branch points $u=0,1/2,\frac{1+\tau}{2},\frac{\tau}{2}$, which may be chosen to map to $0,x,1,\infty$ respectively. A M\"obius map fixes three of these, and then $x$ is determined in terms of $\tau$ as the modular $\lambda$ function $x=\lambda(\tau)$ as in \cref{sec:crossingmodular}.

\begin{figure}
	\centering
	\begin{subfigure}[b]{.55\textwidth}
	\begin{picture}(200,100)
		\put(5,10){\includegraphics[width=.9\textwidth]{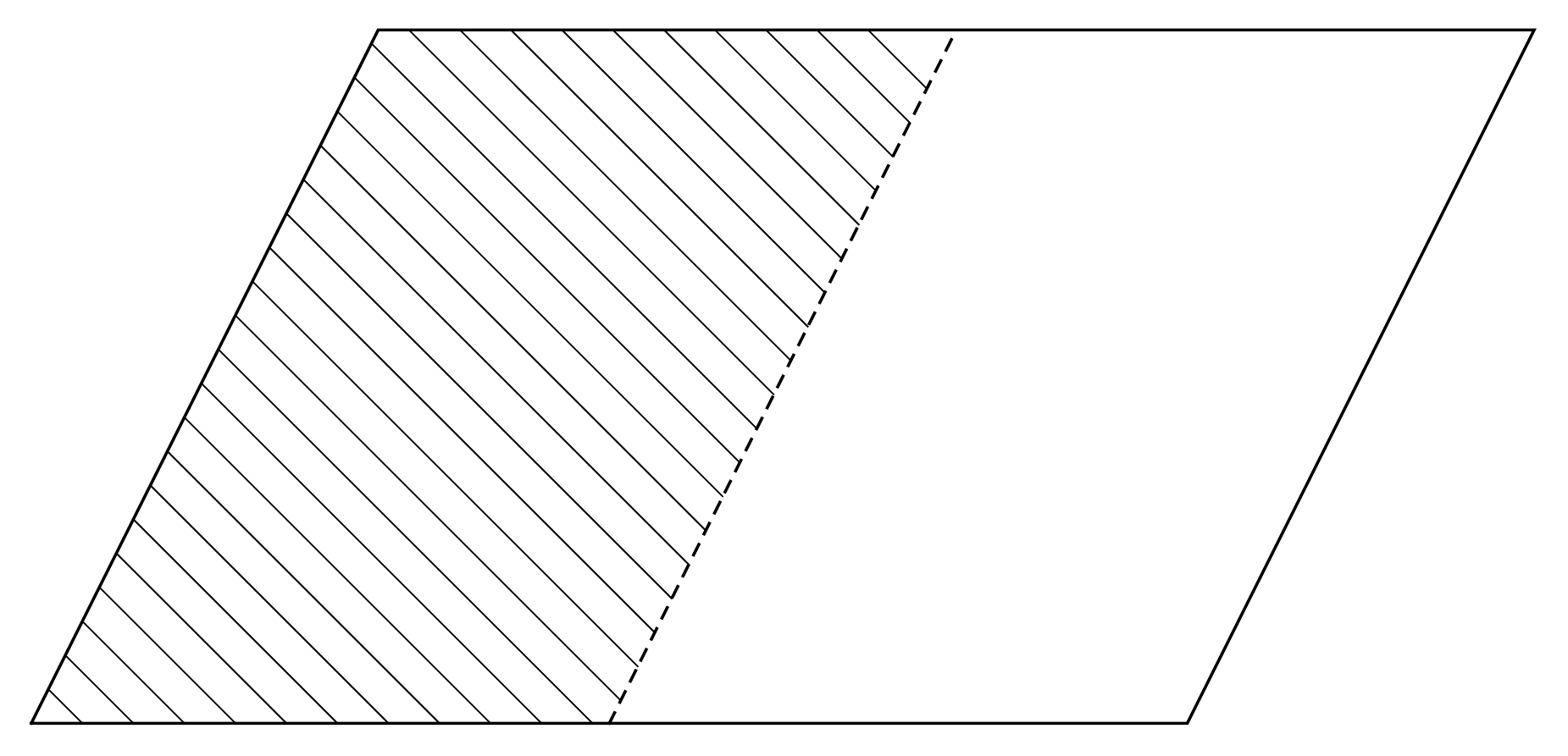}}
		\put(0,2){$0$}
		\put(85,2){$\frac{1}{2}$}
		\put(170,2){$1$}
%		\put(25,65){$\frac{\tau}{2}$}
%		\put(100,65){$\frac{1+\tau}{2}$}
%		\put(198,62){$1+\frac{\tau}{2}$}
		\put(125,118){$\frac{1}{2}+\tau$}
		\put(50,118){$\tau$}
		\put(210,118){$1+\tau$}
	\end{picture}
	\caption{$u$-plane}
	\end{subfigure}
	\begin{subfigure}[b]{.4\textwidth}
	\begin{picture}(100,100)
		\put(15,0){\includegraphics[width=.8\textwidth]{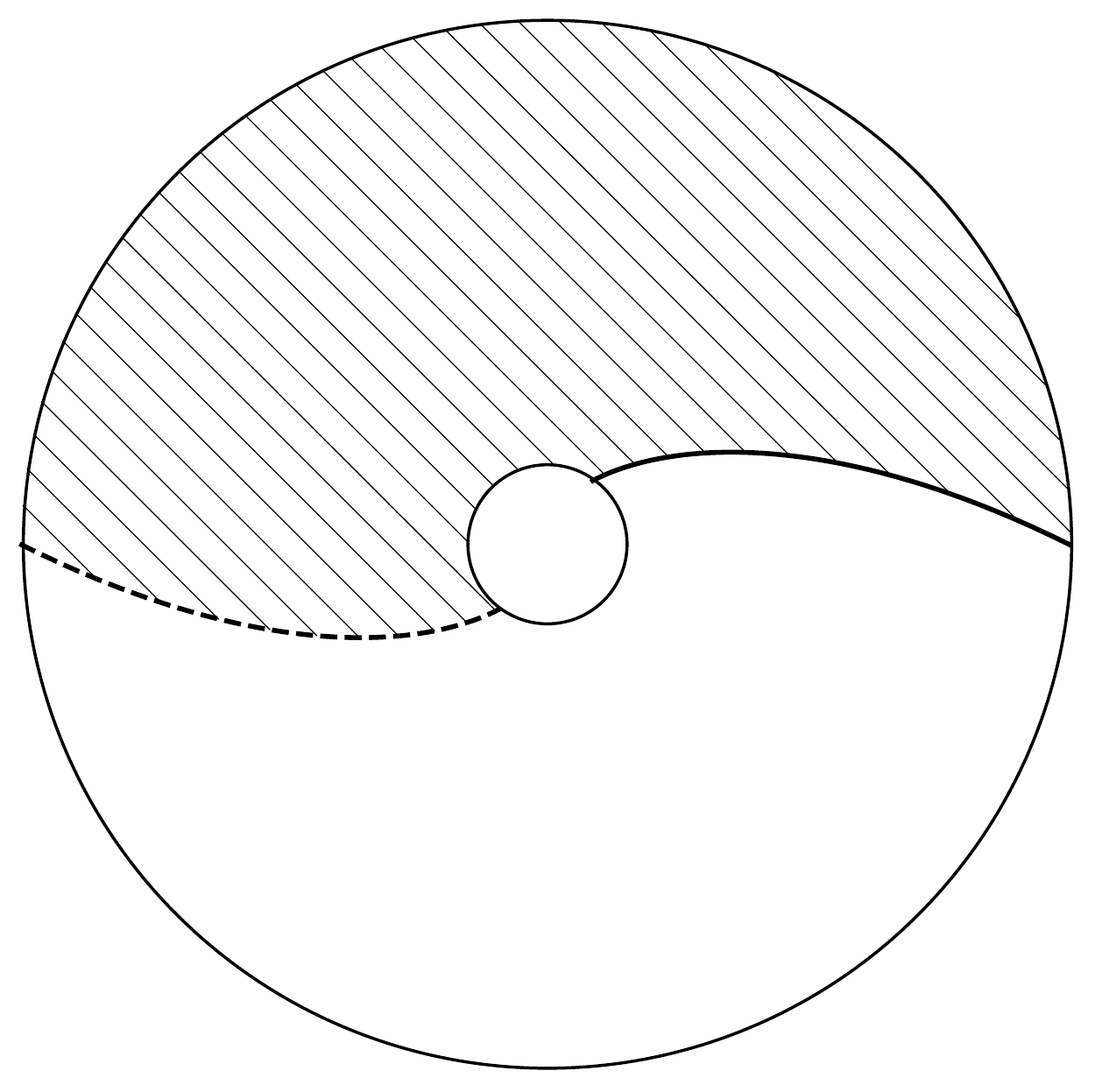}}
		\put(88,84){$q$}
		\put(70,53){$-q$}
		\put(157,67){$1$}
		\put(0,67){$-1$}
	\end{picture}
	\caption{$w$-plane}
	\end{subfigure}
	\caption{The fundamental domain for the plane (hatched) and its double cover, the torus, in the $u$ and $w$ planes. For the torus, the top/bottom and left/right edges of the diamond in the $u$-plane are identified, and the inner and outer circles in the $w$-plane (with a twist). The $\ZZ_2$ further identifies the hatched and unhatched regions.\label{fig:uAndwRegions}}
\end{figure}
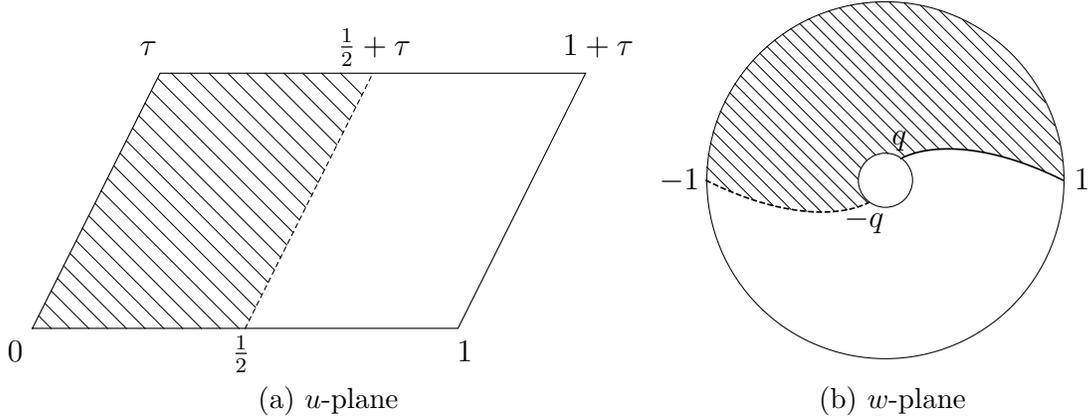

To describe the bulk solutions, it is convenient to write the boundary in terms of the coordinate $w=\exp(2\pi i u)$, which implements the identification $u\sim u+1$ automatically. The other identification to obtain the torus becomes $w\sim q w$, with $q=e^{2\pi i \tau}$, and the $\ZZ_2$ identification giving the plane is $w\sim 1/w$, with fixed points at $w=\pm 1$ and $w=\pm q^{1/2}$. The fundamental domain for these identifications in the $u$ and $w$ coordinates is shown in \cref{fig:uAndwRegions}.

Now take the upper half-space model of Lobachevsky space, with coordinates $w\in\CC$ and $y>0$, and metric
\begin{equation}\label{hypMetric}
	ds^2 = \frac{dw d\bar{w}+dy^2}{y^2}.
\end{equation}
We now may quotient the bulk by isometries which restrict on the boundary to the identifications described above. Firstly, identifying by $(w,y)\sim (q w, |q|y)$ results in a solid torus, with smooth hyperbolic metric since this map acts without fixed points. This construction gives every such metric with torus boundary (and without cusps). Then the $\ZZ_2$ to return to the sphere on the boundary extends isometrically into the bulk as $(w,y)\sim \frac{1}{w\bar{w}+y^2}(\bar{w},y)$, resulting in the ball with conical defects along two curves. These defects appear along the fixed points of the isometry, which are the semicircles $|w|^2+y^2=1$ with $w$ real, and $|w|^2+y^2=|q|$ with $q^{-1/2}w$ real.

Now, solutions with $q$ related by a modular transformation (modulo the $\ZZ$ subgroup generated by $\tau\mapsto\tau+1$ leaving $q$ invariant) have a torus with the same conformal structure on the boundary, but different topology in the bulk: in terms of the rational number used to classify tangles, the cycle described in the $u$-plane by a line through $u=0$ and $u=r+\tau$ (or $u=1$ if $r=\infty$) is contractible in the bulk. Here, $r$ is rational so that this line intersects a lattice point, to form a closed cycle. In the quotient, the fact that the conical defects have the topology of rational tangles follows since the bulk is continuously deformed by moving through $\tau$-space, and the conical defects never intersect.

Finding the on-shell action is easy because of a fortuitous cancellation: the conical defect in the geometry means that there is a delta-function in the curvature supported on the particle worldline, contributing a piece to the Einstein-Hilbert action proportional to the length of the worldline $L$, but this is precisely cancelled by the particle action $mL$ itself. As a consequence, we need only compute the usual Einstein-Hilbert action away from the defect, without taking the singular piece into account. This is particularly useful in the current context, since it means we may compute the action by passing to the smooth double cover, use existing results, and simply halve that action to find our answer.

We therefore need only the solid torus action, regulated according to the boundary metric $ds^2=dzd\bar{z}$ (modified at the operator insertions to regulate the conformal factor between the plane and torus, justified by requiring that the two-point functions are canonically normalised). The action in the flat $dud\bar{u}$ metric on the torus is straightforward to compute, and to convert this to the required $dzd\bar{z}$ metric we need a factor from the conformal anomaly, much as in the twist operator correlation function calculation \cite{Lunin:2000yv} or the transformation between flat metric and `pillow metric' operators described in \cite{Maldacena:2015iua}. The saddle-point contribution $e^{-S}$ in the end factorises into a holomorphic times (conjugate) antiholomorphic piece, the holomorphic half giving the semiclassical block\footnote{Note that the convention used in this section for conformal blocks differs from that in \cref{sec:poincaresum} by a factor of $(1-x)^{(h_{tot}/3)-h_2-h_3}x^{(h_{tot}/3)-h_1-h_2}$ (where $h_{tot}=\sum_i h_i$).}:
\begin{align}\label{c32block}
	\mathcal{F}(c,0,h=c/32;x)&\sim(2^8 x(1-x))^{-c/48} \exp\left(\frac{2\pi c}{48}\frac{K(1-x)}{K(x)}\right)\\
	&= \left(4\frac{\theta_2(q)\theta_4(q)}{\theta_3(q)^2}\right)^{-c/12} q^{-c/48}.
\end{align}
Here $K$ is the elliptic integral, which has branch cuts; the expression in the second line in terms of the upper half-plane parameter $\tau=\frac{1}{2\pi i}\log q$ does not suffer from this ambiguity (excepting possibly for the overall phase from a fractional power).
The first factor in this result comes from the conformal anomaly, and the second factor from the action in the $dud\bar{u}$ metric, $S=-\frac{c}{12}2\pi \Im\tau$, with $\tau=i \frac{K(1-x)}{K(x)}$.

In fact, we can straightforwardly derive a more general solution than this, describing the same external operators, but instead of the internal primary being the identity, we have the exchange of some heavy primary of arbitrary dimension $h_p=\bar{h}_p<c/24$. This means we have two cubic vertices in the bulk, one on each of the original defect worldlines, and, joining the two, the worldline of the intermediate particle. Since this intermediate particle is also heavy, it too sources a conical defect, of arbitrary strength determined by the particle mass. The trick of taking the double cover branched along the worldlines still works, except now the resulting solid torus is not smooth, but has a conical defect determined by the exchanged particle wrapping the nontrivial cycle. To include this, simply alter the hyperbolic bulk metric \cref{hypMetric} by including a defect of the appropriate strength along the line $w=\bar{w}=0$
\begin{equation}\label{conicalMetric}
	ds^2 = \frac{\alpha_p^2}{y^2}\left[\left(\frac{y}{|w|}\right)^{2(1-\alpha_p)}\!\! dw d\bar{w}+dy^2\right],
\end{equation}
and take the same identifications as before.

It is straightforward to generalize the classical action calculation to this case. The simplest way to do this is by differentiating the on-shell action with respect to the mass of the internally exchanged particle. When we differentiate, there is a contribution coming from the variation of the metric and other fields themselves, since the classical solution changes as the mass is changed, but this vanishes because the solution is a stationary point of the action. This leaves only a contribution coming from the explicit variation of the parameter appearing in the action, in this case giving $\frac{dS_\text{on-shell}}{dm}=L$, where $L$ is the length of the worldline of the exchanged particle\footnote{This is true in general, but particularly useful here, as the worldline in question does not end on the boundary, so we do not need to regulate the length.}. In particular, this is why the action reduces simply to worldline length in the limit where $\frac{h}{c}\ll 1$, as in \cite{Hijano:2015qja}, for instance.

 In the metric \eqref{conicalMetric}, this worldline runs along $|w|=0$, between $y=1$ and $y=|q|^{\alpha/2}$ (where the lines of fixed points of the quotient meet the defect at $w=0$), giving $L=\pi\alpha\Im\tau$. Integrating this to find the action, the result in the end matches the one found from the Zamalodchikov monodromy method for semiclassical conformal blocks \cref{hpExchangeblock} (up to the normalisation, discussed in \cref{DOZZ}), which we now briefly describe.

\subsubsection{Monodromy method for semiclassical blocks}\label{monodromy}

A commonly used method for computing semiclassical conformal blocks is the Zamalodchikov monodromy method \cite{Zamolodchikov1987conformal}, reviewed in \cite{Harlow:2011ny,Hartman:2013mia}, which can be understood as coming from the semiclassical limit of Liouville theory. This is essentially equivalent to classical gravity, but since the calculations are, to immediate appearances, rather different, it is nonetheless instructive to include both. It is also a novel example where the monodromy problem can be solved exactly, without any approximations (beyond the semiclassical limit required for its applicability).

Consider the conformal block of four external operators of dimension $h_i=\epsilon_i c/6$, exchanging an operator of dimension $h_p=\epsilon_p c/6$, in the limit $c\to\infty$, with the $\epsilon$'s fixed. To leading order in this semiclassical limit, the block exponentiates as\begin{equation}
	\mathcal{F}(c,h_p,h_i,x)\sim \exp\left(-\frac{c}{6}f(\epsilon_p,\epsilon_i,x)\right).
\end{equation}
The function $f$ is found by solving the differential equation
\begin{equation}
	\psi''(z)+T_c(z)\psi(z)=0
\end{equation}
where $T_c(z)$ is given in terms of one unknown function of $x$, the accessory parameter $c_2(x)$:
\begin{equation}\label{eqq}
	T_c(z)=\left(\frac{\epsilon_1}{z^2} + \frac{\epsilon_2}{(z-x)^2}+ \frac{\epsilon_3}{(z-1)^2}+ \frac{\epsilon_4-\epsilon_1-\epsilon_2-\epsilon_3}{z(z-1)}\right)+\frac{x(1-x)c_2(x)}{z(z-x)(z-1)}.
\end{equation}
As a second order equation, there are two solutions to \rref{eqq}.  These solutions mix when we transport the solution around any topologically non-trivial cycle in the $z$-plane, i.e. we go around any of the singular points of the differential equation.  
 This mixing is described by a monodromy matrix $M$, which has unit determinant by the constancy of the Wronskian. The basis independent data of this matrix is then encoded in the trace of the monodromy matrix.  We then fix the accessory parameter $c_2(x)$ by choosing a particular cycle in the $z$-plane and demanding that the associated Monodromy matrix has
\begin{equation}
	\Tr M= -2\cos \pi\alpha_p,\quad \text{where  } h_p=\frac{c}{24}\left(1-\alpha_p^2\right)
\end{equation}
so the eigenvalues of $M$ are $-e^{\pm i\pi\alpha_p}$.
The conformal block is determined by $c_2(x)=\frac{\partial f}{\partial x}$, the constant of integration determined by normalization (which can be fixed by the behavior as operators become coincident).  The choice of cycle in the differential equation determines the channel of the block.

In the present case, we have $\epsilon_i=3/16$ and the ODE is solved by
\begin{equation}\label{tprime}
	\psi_{\pm}(z) = \frac{1}{\sqrt{t'(z)}}e^{\pm i k t(z)},\quad \text{with }t'(z)=\frac{1}{\sqrt{z(z-x)(z-1)}}.
\end{equation}
The accessory parameter is $c_2(x) = \frac{1-2x+8k^2}{8x(1-x)}$. This is the WKB solution used to work out the limit of large internal dimension \cite{Zamolodchikov1987conformal}, but for the correct accessory parameter, with these values for the external dimensions, it is in fact an exact solution.

 If we choose a cycle enclosing $0$ and $x$ the monodromy is diagonal in this basis.  In particular, the solutions pick up factors of $-\exp({\pm 2 i k t(x)})$: the sign comes from the square root in the prefactor, since $t'(z)$ winds once round the origin as we traverse the cycle, and the phase comes from integrating $t'(z)$ in \cref{tprime} around the cycle from zero to $x$ and back again.
Expressing $t(x)$ as the elliptic integral $2K(x)$ (with appropriate branch choice) we use the monodromy condition to find
\begin{equation}
	c_2(x)=\frac{1-2x}{8x(1-x)} + \frac{\pi^2\alpha^2}{16x(1-x)K(x)^2}
\end{equation}
where $\alpha=\sqrt{1-4\epsilon_p}$. 
This gives the block
\begin{equation}
\label{hpExchangeblock}
	\mathcal{F}(c,h_p,h_i=c/32;x)\sim 2^{4h_p}(2^8x(1-x))^{-c/48} \exp\left(\left(\frac{c}{24}-h_p\right)\pi\frac{K(1-x)}{K(x)}\right).
\end{equation}
This reduces to \cref{c32block} in the case $h_p=0$, with the additional factors coming from the worldline action of the exchanged particle, as discussed above.
In terms of $q=e^{2\pi i \tau}$ the block is
\begin{equation}
	\mathcal{F}(c,h_p,h_i=c/32;x)\sim 2^{4h_p}\left(4\frac{\theta_2(q)\theta_4(q)}{\theta_3(q)^2}\right)^{-c/12} q^{h_p/2-c/48}.
	\end{equation}

These semiclassical blocks give the classical contribution to the correlation function coming from individual saddle points. To find the full correlation function, we should sum over all saddle points, which come from taking $\tau$ to one of its modular images. Thus, gravity naturally leads us to the conformal block Farey tail. It is natural to ask now what the full CFT operator content and couplings are that give a correlation function of this form, but we leave this question for the future.

\subsubsection{Worldline interpretation of heavy exchange, OPE coefficients, and the semiclassical DOZZ formula\label{DOZZ}}

As discussed in \cref{gravitySols}, the modification of the block when we include a heavy internal operator exchange can be understood from the worldline quantized gravity point of view.  The change in the action from including the additional defect accounts for the factor of $\exp\left(-h_p \pi \frac{K(1-x)}{K(x)}\right)$ in the block. The prefactor $2^{4h_p}$, that we fixed by the $x\to0$ limit, does not appear in the gravitational action. This is because the saddle point action computes not just the (holomorphic times antiholomorphic) block, but the contribution of the block to the correlation function, which includes (the leading semiclassical part of) the OPE coefficients $C_{\op\op h_p}^2$.

To find these OPE coefficients, we may compute a three-point function, with a gravitational saddle point consisting of three conical defects from the boundary meeting at a trivalent vertex in the bulk, equivalent to a Liouville theory calculation giving the semiclassical limit of the DOZZ formula \cite{Dorn:1994xn,Zamolodchikov:1995aa,Harlow:2011ny}. In the case of interest, when $\op$ is the $h=c/32$ scalar, the relevant OPE coefficient is $C_{\op\op h_p}=2^{-4h_p}$, cancelling precisely the prefactor in the block. This can be shown directly in this special case by performing the gravity calculation using a double cover trick similarly to before, but also can be obtained from the more general (though much more complicated) results on the semiclassical DOZZ formula, as we now briefly show.

When properly normalized, the OPE coefficients of heavy $h<c/24$ scalar operators are given by $\exp\mathcal{P}(\eta_1,\eta_2,\eta_3)$, where the dimensions of the operators are $h_i=\frac{c}{6}\eta_i(1-\eta_i)$ with $0<\eta<1/2$, and the function $\mathcal{P}$ is \cite{Chang:2016ftb}
\begin{align}
	\mathcal{P}(\eta_1,\eta_2,\eta_3)& \\
	= \frac{c}{6}\Big[&F(2\eta_1)-F(\eta_2+\eta_3-\eta_1) +(1-2\eta_1)\log(1-2\eta_1)+ (\text{2 permutations})\nonumber\\
	&+F(0)-F(\eta_1+\eta_2+\eta_3)-2(1-\eta_1-\eta_2-\eta_3)\log(1-\eta_1-\eta_2-\eta_3)\Big]\nonumber
	\end{align}
	where
	\be
F(\eta)=\int_{1/2}^\eta \log\frac{\Gamma(x)}{\Gamma(1-x)}dx,\quad\text{ for }0<\eta<1\nonumber
\ee
is, roughly speaking, the semiclassical limit of $\Upsilon_b$ which appears in the general DOZZ formula. Taking the case of interest, for which $\eta_1$ is arbitrary and $\eta_2=\eta_3=1/4$, we find
\begin{equation*}
	\mathcal{P}\left(\eta,\frac{1}{4},\frac{1}{4}\right)=\frac{c}{6}\left[F(2\eta)-F\left(\frac{1}{2}-\eta\right)-F\left(\frac{1}{2}+\eta\right)-2F(\eta)+F(0)-2\eta\log2\right].
\end{equation*} To simplify this expression, it is easiest to first differentiate, getting
\begin{equation*}
	\frac{d}{d\eta}\mathcal{P}\left(\eta,\frac{1}{4},\frac{1}{4}\right)=\frac{c}{3}\log\left[\frac{\Gamma(\frac{1}{2}-\eta)\Gamma(1-\eta)\Gamma(2\eta)}{2\Gamma(\frac{1}{2}+\eta)\Gamma(\eta)\Gamma(1-2\eta)}\right] = \frac{c}{3}\log 2^{4\eta-2}
\end{equation*}
where the last equality uses the duplication identity $\Gamma(z)\Gamma(z+\frac{1}{2})=2^{1-2z}\sqrt{\pi}\Gamma(2-z)$ once on the top and once on the bottom. Integrating, and fixing the constant by noting $\mathcal{P}\left(0,\frac{1}{4},\frac{1}{4}\right)=0$ as follows from canonical normalization of the operators, we at last find that $\mathcal{P}\left(\eta,\frac{1}{4},\frac{1}{4}\right)=-4h_p \log2$, reproducing the OPE coefficient claimed above.

\subsubsection{Connection to twist operator correlation functions and the black hole Farey tail}

Finally, let us briefly expand on the connection between these calculations and the four-point function of twist operators in a $\ZZ_2$ orbifold theory, or equivalently the second R\'enyi entropy of two intervals. Firstly, to be clear, we do not demand that the our theory is a $\ZZ_2$ orbifold theory, or that the $h=c/32$ scalar a twist operator; we only want an operator of this dimension so we can use the convenient trick to find classical solutions, and do not necessarily want, for example, the additional light states that must appear in an orbifold CFT\footnote{This is different, for example, to the discussion of \cite{Balasubramanian:2014sra}, which requires the defect geometry to be dual to be a twisted sector state in an orbifold theory.}. An example of an orbifold theory containing twist operators with a gravitational dual is given by the D1-D5 system at the orbifold point, though this is very `stringy' and the low-energy physics bears little resemblance to Einstein gravity. Having said all this, since the conformal block is a universal kinematical function, we may derive it using any theory and operator we like with the correct central charge and dimensions, including an orbifold theory and twist operators.

The conformal block for a given internal primary operator can be defined as the correlation function, with the insertion of a projector onto the descendant states of that primary on a cycle separating the points $0,x$ from $1,\infty$. Taking the external operators as $\ZZ_2$ twist operators, when we pass to the covering space this projection is on a nontrivial cycle of the torus, so we project onto a subset of states propagating round a complementary cycle. It is therefore tempting to use this to identify the conformal block with external weights $c/32$ with a Virasoro character. But this is not quite right: the projection to obtain the Virasoro character leaves more states intact, because it contains not just descendants in the orbifold theory, but also all descendants in the seed theory, which includes states regarded as Virasoro primaries from the $\ZZ_2$ orbifold theory. One way of saying this is that the untwisted sector of the orbifold theory (relevant since all states exchanged are untwisted) has an extended algebra, the symmetric product of two Virasoros, one from each copy of the theory; the character includes descendants under this entire algebra, but the block only descendants under the diagonal Virasoro. The character and the block do (when the appropriate conformal anomaly is included) match in the semiclassical limit, but not the perturbative corrections. 
 The OPE coefficient $2^{-4h_p}$ that appears from the gravitational calculation also matches the coefficient between two twist operators and a third primary operator to which they fuse (which must be untwisted, and of the form $\phi^{(1)}\otimes\phi^{(2)}$, where $\phi$ is some primary in the seed theory and the superscript indicates which copy it acts on).
 
 Finally, we directly connect to previous work by noting that the black hole Farey tail is a special case of our conformal block Farey tail, where the CFT is taken to be a $Z_2$ orbifold of a gravity theory and we consider the correlation function of twist operators, since this is (up to an anomaly term) just the partition function in the original theory \cite{Lunin:2000yv}.

\subsubsection{Blocks computed perturbatively in $h/c$.}
As mentioned already, there is a convenient limit in which to study semiclassical blocks, where the dimension of some operator is large, but much less than $c$. Concretely, one may solve the monodromy problem described in \cref{monodromy} perturbatively in $\epsilon=6h/c$ for the appropriate operator. To leading order, as discussed above this corresponds to a `probe limit' in gravity, where the worldlines of the operator in question become geodesics in the background created by other operators.

One might try to apply the ideas discussed in this section to this perturbative limit, for example for the four-point function $\langle \op_H \op_H \op_h \op_h \rangle$, where we compute exactly in the dimension $H$ (of order $c$) and perturbatively in $h/c$. In this example, the perturbation theory describes a conical defect background created by $\op_H$, and a geodesic associated with $\op_h$ in this background. However, in this geodesic limit, one runs into trouble when attempting to analytically continue in the cross-ratio. In particular, along some curve (depending on the dimension $H$) in $\tau$ space, the geodesic intersects the defect, and a na\"ive analytic continuation of the block past that curve gives results that are not reproduced by any geodesic. From the gravity point of view, there is no reason why analytic continuation should be applicable, since the spacetime is not analytic. Nonetheless there is no obvious breakdown in perturbation theory from the point of view of the monodromy method, so it is likely that this tension can be resolved only by going beyond perturbation theory.

From the gravitational point of view, it is natural that the perturbation theory ceases to be applicable when the geodesic intersects the conical defect: once the worldlines are parametrically close, it is not valid to neglect their mutual gravitational interaction. This interaction may prevent the worldlines from crossing, in which case the topological discussion of rational tangles remains applicable, though the nontrivial tangling of the worldlines may be confined to a parametrically small region of the spacetime.

This question would benefit from more quantitative understanding, particularly as it is an important limit for holographic calculations of entanglement entropy. Another example in this spirit, where progress may be easiest, is for the four-point function of identical operators $\langle \op_h \op_h \op_h \op_h \rangle$, as considered in \cite{Hartman:2013mia}, relevant for the entanglement entropy of two disjoint intervals. The problem there occurs when the cross-ratio hits the line $\Im(x)=0,\Re(x)>1$, where two geodesics intersect. Na\"ive perturbation theory suggests that the geodesics pass through one another, so the conformal blocks are single valued in cross-ratio space, but this seems incompatible with our results at finite $\epsilon$.

\section*{Acknowledgements}

We are grateful to V. Balasubramanian, A. Bernamonti, M. Cheng, F. Galli, T. Gannon, T. Hartman, C. Keller, E. Perlmutter and D. Poland for useful conversations. A.M.~is supported by the National Science and Engineering Council of Canada and by the Simons Foundation. H.M.~is supported by a fellowship from the Simons Foundation. G.N.~is supported by an NSERC Discovery Grant.

\appendix

\section*{Appendix: conformal blocks and $PSL(2,\ZZ)$ representations in minimal models}
In this appendix, we review the Coulomb-gas representation of the conformal blocks for minimal models, from which we obtain the representation of the modular group associated with various conformal blocks. Our discussion is based mainly on \cite{francesco2012conformal,Dotsenko:1985hi,Dotsenko:1984ad,Dotsenko:1984nm}, with some slightly different conventions, more convenient for our purposes.
For simplicity, we will focus here on the correlation functions of identical second-order scalar operators.
The extension to mixed correlators or operators with spin can be found in \cite{francesco2012conformal,Dotsenko:1985hi,Dotsenko:1984ad,Dotsenko:1984nm}.

The four-point function of the scalar operator $\phi_{(r,s)}$ of dimensions $h=\bar{h}=h_{(r,s)}$
\begin{equation}
\langle \phi_{(r,s)}(z_1)  \phi_{(r,s)}(z_2)  \phi_{(r,s)}(z_3)  \phi_{(r,s)}(z_4)\rangle
= \left|z_{12}  z_{34}
\right|^{-4h_{(r,s)}}
\left|
\frac{(1-x)}{x^2}
\right|^{-4h_{(r,s)}/3} G_{r,s}(x,\xbar)\,,
\end{equation}
decomposes into conformal blocks as
\begin{equation}
G_{r,s}(x,\bar{x}) = \sum_{i=1}^{N_{(r,s)}} C^2_{(r,s)(r,s)(r_i,s_i)} \calF_i(x) \bar{\calF}_i(\bar{x})
\end{equation}
with $N_{(r,s)}$ primary fields appearing in the fusion rule
\begin{equation}
\phi_{(r,s)}\times \phi_{(r,s)} =\sum_{i=1}^{N_{(r,s)}} \phi_{(r_i,s_i)}\,.
\end{equation}
The conformal block $\calF_i$ is associated with the primary $\phi_{(p_i,q_i)}$ appearing in the fusion rule. We shall organize the label $i$ such that $h_{(r_i,s_i)}$ is a non-decreasing function of $i$, with $h_{(r_1,s_1)}=0$ (i.e.~$\phi_{(r_1,s_1)}=\id$). This definition of $\calF_i$ implies the leading order behaviour
\be\label{eq:calFnorm}
\calF_i(x) = x^{h_{(r_i,s_i)}-4h_{(r,s)}/3}+\cdots
\end{equation}
as $x\rightarrow 0$.

In the Coulomb-gas formalism (see section 9.2.3 of \cite{francesco2012conformal}), the holomorphic four-point function of $\phi_{(r,s)}$ is computed by
\begin{align}
\langle \phi_{(r,s)}(z_1)\phi_{(r,s)}(z_2)&\phi_{(r,s)}(z_3)\phi_{(r,s)}(z_4)\rangle \nonumber \\
	&=	\langle V_{(r,s)}(z_1) V_{(r,s)}(z_2) V_{(r,s)}(z_3) V_{(-r,-s)}(z_4) Q^{r-1}_+ Q^{s-1}_-
\rangle\,.
\end{align}
The screening operator $Q_{\pm}$ is defined by
\begin{equation}
Q_{\pm} \equiv \oint_C dw ~V_{ \alpha_{\pm}}(w)
\end{equation}
with $\alpha_\pm\equiv \alpha_0 \pm \sqrt{\alpha_0^2+1}$ and $\alpha_0\equiv 1/(2\sqrt{p(p-1)})$, or equivalently $\alpha_\pm \equiv  \pm \left[\frac{p}{p-1}\right]^{\pm\frac{1}{2}}$. The contour $C$ must be chosen appropriately to get the full correlation function; a different choice of contour will give the contribution from an individual conformal block as we will see.

Let us now focus on the case of $(r,s)=(2,1)$\footnote{The results for $(r,s)=(1,2)$ can be obtained straightforwardly from this by taking $p\to 1-p$ in the final expressions}, with dimension
\begin{equation}
h_{(2,1)}=\frac{p+2}{4 (p-1)}.
\end{equation} For example for $(p,p')=(4,3)$ we have $c=1/2$ with $h_{2,1}=1/2$, and for $(p,p')=(5,4)$, $c=7/10$ and $h_{2,1}=7/16$, giving the scalar operators usually labelled $\epsilon$ in the Ising model and $\sigma'$ in the tricritical Ising model.
We also have
\begin{equation}
\alpha_+=\sqrt{\frac{p}{p-1}},\quad
\alpha_{2,1}=-\frac{1}{2 }\sqrt{\frac{p}{p-1}}\,.
\end{equation}

The fusion rule of two $\phi_{(2,1)}$ operators is given by
\begin{equation}
\phi_{(2,1)}\times \phi_{(2,1)}=\phi_{(1,1)}+\phi_{(3,1)}
=\id+\phi_{(3,1)},
\end{equation}
the dimension of the $\phi_{(3,1)}$ operator given by $h_{3,1}=(1 + p)/(-1 + p)=-(2a+1)$, where we have introduced the parameter $a=2\alpha_+ \alpha_{2,1}=-\frac{p}{p-1}$.
We will find the conformal blocks for the four-point function of $\phi_{(2,1)}$ with these two operators exchanged, in terms of hypergeometric functions.

From the Coulomb gas expressions, the four-point function is given by
\begin{align}
	\langle \phi_{(2,1)}(z_1)\phi_{(2,1)}(z_2)&\phi_{(2,1)}(z_3)\phi_{(2,1)}(z_4)\rangle \nonumber\\
	&=\oint_C dw\; \langle  V_{(2,1)}(z_1) V_{(2,1)}(z_2) V_{(2,1)}(z_3) V_{(-2,-1)}(z_4) V_{+}(w) \rangle\,,
\end{align}
and using the formula for the $k$-point function of vertex operators
\begin{equation}
\langle V_{\alpha_1}(z_1)\ldots V_{\alpha_k}(z_k)\rangle
= \prod_{i<j} (z_{ij})^{2\alpha_i \alpha_j},\quad \text{with}\quad z_{i,j}\equiv z_i-z_j
\end{equation}
gives the integral expression
\begin{equation}
\langle \phi_{(2,1)}(0)\phi_{(2,1)}(x)\phi_{(2,1)}(1)\phi_{(2,1)}(\infty)\rangle =\left[(1-z) z\right]^{2 \alpha _{2,1}^2}
\oint_C dw~
\left[ w (w-1)  (w-x) \right]^a\,.
\end{equation}

The conformal blocks can be extracted from this expression simply by changing the contour of integration $C$, as
\begin{equation}
\calF_i (x) = 
 \frac{1}{N_i}
  \left[x(1-x)\right]^{-\left(a+\frac{1}{3}\right)}
   \int_{C_i} dw~
\left[ w (w-1)  (w-x) \right]^a
\end{equation}
with $C_i$ being the line from $0$ to $x$ for the vacuum block, and from $1$ to $\infty$ for the $\phi_{(3,1)}$ exchange. The normalisation is fixed by the \cref{eq:calFnorm}, to give
\begin{equation}
N_1=\frac{\Gamma^2(a+1)}{\Gamma(2a+2)},\quad
N_2=\frac{\Gamma(-3a-1)\Gamma(a+1)}{\Gamma(-2a)}\,
\end{equation}
and the blocks can then be expressed in terms of hypergeometric functions as
\begin{align}
\calF_\id(x)&=
 x^{\frac{2}{3}+a}(1-x)^{-\frac{1}{3}-a}\,\twoFone\left(-a,a+1,2a+2,x\right)=x^{\frac{2}{3}+a}+\cdots\,,\nonumber\\
\calF_{(3,1)}(x)&=\left[x(1-x)\right]^{-\left(a+\frac{1}{3}\right)}\,\twoFone\left(-a,-3a-1,-2a,x\right)=x^{-\frac{1}{3}-a}+\cdots \,. 
\end{align}
With these expressions in hand, the $T$ matrix and the $S$ matrix can be read off by using standard identities for hypergeometric functions, which my be derived by deforming the contour of integration (as in figure 9.3 of \cite{francesco2012conformal}). For example, for the $S$ matrix, we use
\begin{align}
	\calF_\id(1-x)&=-\frac{1}{2\cos(a\pi)}\left(\calF_\id(x)+(1+2 \cos (2 \pi a)) \frac{N_2}{N_1}\calF_{(3,1)}(x)\right)  \, , \nonumber\\
\calF_{(3,1)}(1-x)&=- \frac{1}{2\cos(a\pi)}\left(\frac{N_1}{N_2} \calF_\id(x)- \calF_{(3,1)}(x)\right) \, .
\end{align}
The resulting representation of $PSL(2,\ZZ)$, in the basis $\{\calF_\id,\calF_{(3,1)}\}$, is generated by
\begin{equation}
	T = e^{\frac{2}{3}i\pi}\begin{pmatrix}
		e^{a i\pi}&0\\ 0& -e^{-a i\pi}
	\end{pmatrix},\quad
	S = -\frac{1}{2\cos(a\pi)} \begin{pmatrix}
		1 & \frac{N_1}{N_2} \\ (1+2 \cos (2 \pi a))\frac{N_2}{N_1} & -1
	\end{pmatrix}.
\end{equation}
In our conventions, the matrices mean for example, that $\calF_i(1-x)=\sum_j S_{ji}\calF_j(x)$.
It is easy to see here that rescaling the basis (picking $N_1=\sqrt{1+2 \cos (2 \pi a)}N_2$ instead of the choices above) can make the representation unitary, as long as $1+2 \cos (2 \pi a)>0$, as is the case for $p\geq 5$. The marginal case $p=4$ is discussed in \cref{Isingeeee}.

The usual solution to crossing is now simple to obtain. Imposing $T$-invariance on the correlator restricts it to the form
\begin{equation}
G= |\calF_\id|^2+ C_{(3,1)}^2 |\calF_{(3,1)}|^2,
\end{equation}
and then $S$-invariance gives the OPE coefficient as
\begin{equation}
C_{(3,1)}=\pm\sqrt{2 \cos \left(\frac{2 \pi  p}{p-1}\right)+1}\;\frac{ \Gamma \left(\frac{2p+1}{p-1}\right) \Gamma \left(-\frac{2}{p-1}\right)}{\Gamma \left(\frac{2 p}{p-1}\right) \Gamma \left(-\frac{1}{p-1}\right) }\,.
\end{equation}

%%%%%%%%%%%%
\bibliographystyle{utphys}
\bibliography{fareyBlock}

\end{document}